\documentclass[]{article}

\usepackage{amsmath,amssymb,upgreek}
\usepackage{authblk}
\usepackage{caption}
\usepackage{chngcntr}
\usepackage[thinc]{esdiff}
\usepackage{etoolbox}
\usepackage{fullpage}
\usepackage[a4paper, margin={2cm}]{geometry}
\usepackage{graphicx} 
\usepackage[hidelinks]{hyperref}
\usepackage{makecell}
\usepackage[sort&compress,numbers]{natbib}
\usepackage{physics}
\usepackage{siunitx}
\usepackage[caption=false]{subfig}
\usepackage{xcolor}

\setcellgapes{2pt}

\graphicspath{{'../figures/'}}

\title{Spatiotemporal model of cellular mechanotransduction via Rho and YAP}

\author[1]{Javor K. Novev}
\author[1,2]{Mathias L. Heltberg}
\author[1]{Mogens H. Jensen}
\affil[*]{Corresponding author: mhjensen@nbi.ku.dk}

\author[1,$\dagger$]{Amin Doostmohammadi}
\affil[$\dagger$]{Corresponding author: doostmohammadi@nbi.ku.dk}
\affil[1]{Niels Bohr Institute, University of Copenhagen, Blegdamsvej 17, 2100 Copenhagen \O, Denmark}
\affil[2]{Laboratoire de Physique, Ecole Normale Superieure, Rue Lhomond 15, Paris 07505}

\begin{document}
	
	\maketitle
	
	\begin{abstract}
		How cells sense and respond to mechanical stimuli remains an open question. Recent advances have identified the translocation of Yes-associated protein (YAP) between nucleus and cytoplasm as a central mechanism for sensing mechanical forces and regulating mechanotransduction. We formulate a spatiotemporal model of the mechanotransduction signalling pathway that includes coupling of YAP with the cell force-generation machinery through the Rho family of GTPases. Considering the active and inactive forms of a single Rho protein (GTP/GDP-bound) and of YAP (non-phosphorylated/phosphorylated), we study the cross-talk between cell polarization due to active Rho and YAP activation through its nuclear localization.	 \par
		For fixed mechanical stimuli, our model predicts stationary nuclear-to-cytoplasmic YAP ratios consistent with experimental data at varying adhesive cell area.
		We further predict damped and even sustained oscillations in the YAP nuclear-to-cytoplasmic ratio by accounting for recently reported positive and negative YAP-Rho feedback. Extending the framework to time-varying mechanical stimuli that simulate cyclic stretching and compression, we show that the YAP nuclear-to-cytoplasmic ratio's time dependence follows that of the cyclic mechanical stimulus. The model presents one of the first frameworks for understanding spatiotemporal YAP mechanotransduction, providing several predictions of possible YAP localization dynamics, and suggesting new directions for experimental and theoretical studies.
	\end{abstract}
	
	
	\section{Introduction}
	
	In response to an external mechanical cue, cells balance their shape and cytoskeletal structure through internal forces, creating mechanical feedback. In multicellular tissues, intracellular forces are then transmitted to neighbouring cells through cell-cell junctions. Forces acting within cells provide them with a mechanism to maintain and alter their shape, but what instructs a cell to react in a specific way, e.g. divide, die or move, in response to different forces?

	Inside a cell, a plethora of biochemical signalling pathways work together to induce a specific mechanical response~\cite{Vining2017,Cheng2017}. Therefore, to reveal the mechanical basis of mechanotransduction, it is beneficial to seek simpler and more flexible model systems.
	
	In recent years a single transcription factor, the Yes-associated protein (YAP) and its paralogue TAZ~\cite{Panciera2017}, have been identified as the central hub for different signalling pathways and the master regulator of mechanotransduction~\cite{Panciera2017,Vining2017,Dupont2011,Halder2012,Hansen2015,Ladoux2017,Elosegui-Artola2017,Totaro2018,Totaro2018a}. In response to different mechanical stimuli, such as tension or compression, YAP relocates between the cell nucleus and cytoplasm, regulating the cell response by switching between active (nucleus) and inactive (cytoplasmic) states~\cite{Panciera2017}. As such YAP mechanotransduction controls a wide range of cell behaviours such as self-renewal, differentiation, proliferation, stemness, and apoptosis~\cite{Vining2017,Panciera2017,Ladoux2017,Brusatin2018}. At the multicellular level, this protein exerts control on organ size \cite{Panciera2017, Brusatin2018} and is associated with most malignant properties: unrestrained proliferation-cell survival~\cite{Aragona2013}, chemoresistance~\cite{Kim2016,Zanconato2016}, and metastasis~\cite{Zhang2015,Zanconato2016a}.
	
	For years the activation of YAP was considered to be controlled mainly by LATS (large tumour suppressor) proteins through the Hippo signalling pathway:  LATS phosphorylates YAP in the cytoplasm~\cite{Rausch2019}, thus suppressing its nuclear entry and transcriptional activity~\cite{Hansen2015}. However, recent studies have discovered a LATS-independent mechanical pathway of YAP activation~\cite{Dupont2011,Halder2012,Totaro2018}. By culturing single cells on adhesive areas of different sizes, it was shown that YAP relocates to the cell nucleus and becomes activated on larger contact areas where the cell can spread, while rounded cells on smaller adhesive areas showed cytoplasmic and thus inactive YAP~\cite{Dupont2011,Elosegui-Artola2017}. This dependence of YAP activation/deactivation on the contact area is shown to remain unaffected for cells with depleted LATS protein, indicating that mechanical activation is a parallel pathway to Hippo signalling for controlling the YAP activity.
	
	Similarly, it has been shown that the cells can feel the stiffness of the substrate they are moving on through YAP: on soft substrates cells cannot spread and remain rounded, hence YAP is inactive (cytoplasmic), while YAP becomes active when the cells spread on stiff substrates~\cite{Dupont2011,Halder2012,Elosegui-Artola2017}. Interestingly, manipulating the levels of nuclear or cytoplasmic YAP restores the spreading capability. For example, inducing overexpression of YAP in the nucleus leads to cell spreading even on soft substrates and YAP depletion causes rounded cell shapes on stiff substrates~\cite{Dupont2011}. This indicates that YAP not only allows cells to perceive changes in the mechanics of their environment but is also able to generate mechanical feedback that determines cellular behaviour. A similar feedback mechanism is considered as a possible route to cancer development: abnormal stiffening of the extracellular matrix (ECM) leads to overactivation of YAP, which in turn causes hyperproliferation and further stiffening of the ECM through e.g., modulation of collagen synthesis \cite{Panciera2017} and expressing ECM-modifying enzymes \cite{Zanconato2019}, creating a mechanotransduction cycle for cancer cell invasion~\cite{Panciera2017}. YAP has already been shown to be necessary for the reprogramming of normal mammary cells into tumorigenic ones via changes in the local microenvironment~\cite{Panciera2020}. Moreover, since downstream gene expressions and mutations of YAP have not been reported in human cancer, mechanical inputs from abnormal microenvironments are put forward as the prime candidate to induce YAP overactivation in cancer cells~\cite{Panciera2017,Zanconato2016,Zanconato2016a}. Other types of mechanical cues have been reported to induce YAP activation: stretching of confluent epithelial monolayers~\cite{Panciera2017}, changes in the density of cells within a tissue~\cite{Panciera2017, Aragona2013, Hsiao2016}, perturbations in flow in blood vessels~\cite{Panciera2017, Huang2016a}, and direct mechanical stimulation of the nucleus~\cite{Elosegui-Artola2017, Aureille2019}.
	
	Nonetheless, most experimental ~\cite{Vining2017,Panciera2017,Elosegui-Artola2017,Totaro2018,Totaro2018a,Zhang2015,Sakabe2017,Uhler2017,Zhang2018} and computational~\cite{Cheng2017,Sun2016} studies on YAP conducted recently are dominated by exploring molecular signalling pathways to YAP and its downstream chemical and genetic effectors, while overlooking the role of mechanical forces in directing mechanotransduction. Though some existing models attempt to describe the mechanotransduction signalling pathway in considerable detail - see Refs. \cite{Sun2016,Peng2017} - they assume a well-mixed system. What sets our work apart is that we aim to provide a unified description of mechanical stimuli and the biochemical signalling pathway that is activated when the cell receives such stimuli; to our knowledge, our model is also the first one to look into the spatial distribution of YAP. As nuclear localization is required for the activation of YAP \cite{Halder2012}, understanding the spatial variations in the concentrations of the relevant species is crucial, and well-mixed models cannot naturally provide such insight. Thus, our approach is to formulate a spatiotemporal model of the network which contains a minimum number of chemical components, yet describes the essential features of mechanotransduction. \par 
	
	\section{Computational model}
	
	In this section, we describe the main features of our spatiotemporal computational model for mechanotransduction, focussing on the one-dimensional case. The signalling network that regulates YAP involves many components \cite{Panciera2017}, and accounting for all the complex interrelationships between them is impractical. For the sake of simplicity, our model considers a single explicit regulator of YAP - the Rho family GTPases, which are known to be essential for YAP activation \cite{Panciera2017, Halder2012}. Rho proteins are important in the development of cell polarity \cite{Park2007} and to mechanotransduction \cite{Burridge2019}. In their active form, `classical' Rho GTPases bind to GTP \cite{Burridge2019,Heasman2008} and reside at the cell membrane; in contrast, the inactive forms are GDP-bound and diffuse freely through the cytosol \cite{Olofsson1999}. `Atypical' RhoGTPases also exist - these are predominantly GTP-bound and are thought to be regulated through other mechanisms~\cite{Heasman2008}. The main Rho GTPases that affect cell polarity and migration are RhoA, CDC42 and RAC1 \cite{Heasman2008}; however, following the well-established `wave-pinning' model of Mori et al. \cite{Mori2008}, we take a simplified approach that includes only a single GTPase in its active and inactive forms, but still describes the basic features of cell polarization. Rho activity has been shown to be necessary for YAP nuclear localization and transcriptional activity \cite{Dupont2011,Piccolo2013}, but the activity of YAP modified so as to prevent phosphorylation at serine is unaffected by Rho inhibition~\cite{Dupont2019}. Kofler et al.~\cite{Kofler2018} have found that nuclear transport of the closely related YAP paralogue, TAZ, is regulated by RhoA; moreover, they have identified nuclear localization and efflux signals in TAZ that are conserved in YAP.  \par
	
	\subsection{Coupled reaction-diffusion equations for Rho and YAP}
	
\begin{figure}
		\centering
		\includegraphics[width=12cm]{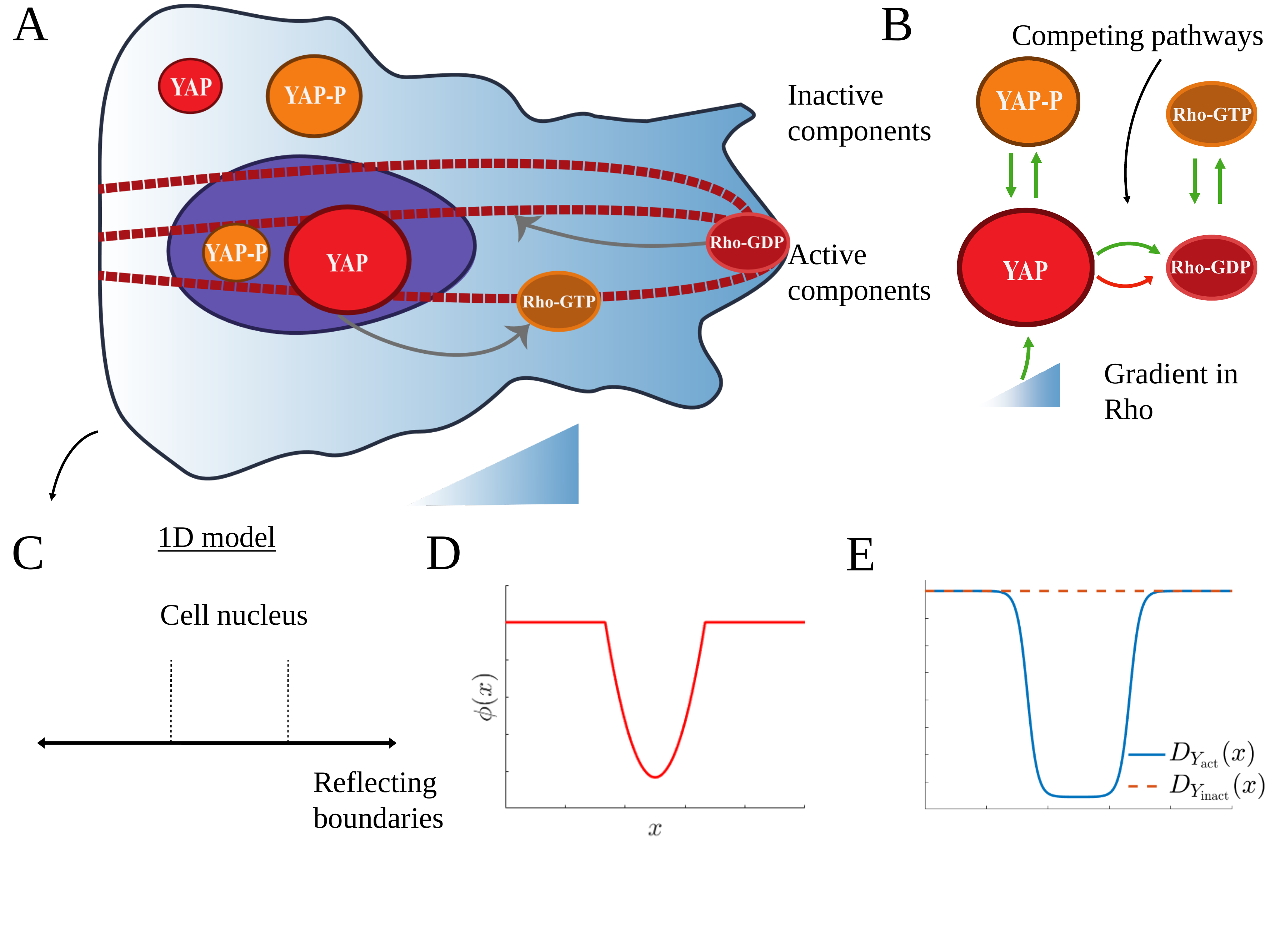}				
		\caption{\textbf{A}. Cartoon of our model for the mechanotransduction in the cell. GTP-bound activated Rho ($A_\mathrm{act}$) localizes to the cell membrane and is both produced from and converted into its cytosolic inactive form ($A_\mathrm{inact}$), which is bound to GDP. This generates a gradient in Rho that activates YAP and through stimulating the formation of actin cables (thick dashed lines) causes its active form ($Y_\mathrm{act}$), which is produced from and degraded to phosphorylated inactive YAP ($Y_\mathrm{inact}$, YAP-P), to localize to the nucleus. In turn, nuclear YAP affects the balance between active and inactive Rho. In addition to participating in chemical reactions, all species diffuse through the cell. \textbf{B}. Schematic of our model for the mechanotransduction signalling pathway illustrating the various feedback mechanisms in the system. \textbf{C}. Mapping of the model to 1D. \textbf{D}. Harmonic potential acting on YAP in the nuclear region. \textbf{E}. Diffusion coefficients for the the two forms of YAP in our 1D representation of the cell. The active form (solid line) binds to TEAD in the nucleus, which corresponds to a drop in its diffusivity in that region, whereas the diffusion coefficient of the inactive form (dashed line) is uniform throughout the cell.}
		\label{fgr:YAP_schematic}
	\end{figure}

	We model the effect of Rho on the interconversion and spatial localization of active and inactive YAP, i.e., YAP phosphorylated at one of its serine residues \cite{Rosenbluh2012, Yu2015,Das2016}. To this end, we formulate the following system of reaction-diffusion equations:
	\begin{equation}\label{eq1}
	\pdv{ A_\mathrm{act}}{t} =	\div{\left(D_{A_\mathrm{act}}\grad{A_\mathrm{act}})\right.} + f_A (A_\mathrm{act}, A_\mathrm{inact}, Y_\mathrm{act});	
	\end{equation}
	
	\begin{equation}\label{eq2}
	\pdv{ A_\mathrm{inact}}{t} =  \div{\left(D_{A_\mathrm{inact}}\grad{A_\mathrm{inact}})\right.} - f_A (A_\mathrm{act}, A_\mathrm{inact}, Y_\mathrm{act});	
	\end{equation}
	
	\begin{equation}\label{eq3}
	\pdv{Y_\mathrm{act}}{t} = -\div{\emph{\textbf{J}}_{Y_\mathrm{act}}} + f_Y(A_\mathrm{act}, Y_\mathrm{act}, Y_\mathrm{inact});
	\end{equation}
	
	\begin{equation}\label{eq4}
	\pdv{ Y_\mathrm{inact}}{t} = -\div{\emph{\textbf{J}}_{Y_\mathrm{inact}}} - f_Y(A_\mathrm{act}, Y_\mathrm{act}, Y_\mathrm{inact}),
	\end{equation}
	
	where $A_\mathrm{i}$ and $Y_\mathrm{i}$ are the concentrations of the different forms of Rho and YAP and $D_{A_\mathrm{i}}$ are the diffusion coefficients of active and inactive Rho; note that $D_{A_\mathrm{inact}} \gg D_{A_\mathrm{act}}$ because the active form is bound to the cell membrane. $f_\mathrm{i}$ denotes the rate of generation of the active form of Rho or YAP, the fluxes of the different YAP species are $\emph{\textbf{J}}_{Y_\mathrm{i}}$, and $t$ is time. We nondimensionalize the concentrations of the different forms of Rho by an arbitrary value $ C_{0 \mathrm{A}}$ so that they correspond to those reported by Mori et al. \cite{Mori2008}; for YAP, we normalize the total amount in  the cell $N_{0 \mathrm{Y}} = \int_{V} \left(Y_\mathrm{act} + Y_\mathrm{in act}\right)\differential V$ to unity, with $V$ being the cell volume. We present a schematic diagram of the model network in Figure~\ref{fgr:YAP_schematic}.
	
	\subsection{Rate of Rho generation}
	
	We base our expression for the rate of generation of active Rho on a well-known model according to which the concentration of active Rho forms polarized stationary profiles (`pinned waves') \cite{Mori2008}:
	\begin{equation}\label{eq5}
	f_A (A_\mathrm{act}, A_\mathrm{inact}, Y_\mathrm{act}) = \left(c_1 + \frac{c_2 A_\mathrm{act}^2}{c_3^2 + A_\mathrm{act}^2} \right)A_\mathrm{inact}-c_4 A_\mathrm{act} + c_5\left(1- \frac{\int_{S_\mathrm{nucl}}Y_\mathrm{act}\differential S} { Y_{\mathrm{act \ nucl \ threshold}}}\right),										
	\end{equation} 			  
	where $c_1-c_4$ are constants (see Table~\ref{tbl:main_parameters} for their units), $c_5$ is measured in $\mathrm{s^{-1}}C_{0 \mathrm{A}}$, and the integral is over the extent of the nucleus $S_\mathrm{nucl}$. The first two terms in eq.~\eqref{eq5}, whose form we take from Ref. \cite{Mori2008},	allow for bistable behaviour of $A_\mathrm{act}$\cite{Mori2008}, for which there is experimental evidence \cite{Byrne2016}. With $c_5 = 0$ and suitable values for the other constants (see Table~\ref{tbl:main_parameters}), a local perturbation to the homogeneous system can generate a concentration profile that takes approximately the value of one stable stationary solution for $A_{\mathrm{act}}$ on one end of the cell, reaches the approximate value of the other stable stationary $A_{\mathrm{act}}$ on the opposite end, and undergoes a steep transition in between (`pinned wave'). This concentration profile is a 1D representation of a polarized cell state. \par
	
	We introduce the last term in eq.~\eqref{eq5} to account for the coupling between Rho and nuclear YAP. This is motivated by recent experiments that demonstrate the existence of both positive feedback of nuclear YAP on Rho via ARHGAP 28 and negative feedback via ARHGAP 29 and NUAK2 \cite{Mason2019}. However, neither the relative magnitudes of these feedbacks are known, nor whether or under what conditions the overall effect of YAP on Rho can change sign. Here we introduce a feedback term that allows for such a change of sign, and consider the different regimes that arise depending on where the system is located in parameter space. To this end, we assume that the localization of YAP to the nucleus activates Rho until the total nuclear YAP exceeds the threshold value $Y_{\mathrm{act \ nucl \ threshold}}$. In the limit $Y_{\mathrm{act \ nucl \ threshold}} \to 1$, YAP can only activate Rho, although the extent of activation decreases as more active YAP is localized to the nucleus; the limit $Y_{\mathrm{act \ nucl \ threshold}} \to \infty$ corresponds to activation due to YAP independent of the amount of active YAP in the nucleus. Since we do not have data on the relative strength of the different feedback loops, we explore different scenarios below. Though in general the rate of activation of Rho by YAP would also be a function of both $A_{\mathrm{act}}$ and $A_{\mathrm{inact}}$, the form we choose for it in the last term of eq.~\eqref{eq5} has the essential features of a system with both positive and negative feedback, yields a broad range of dynamic behaviours, and its simplicity facilitates the analysis and enhances the predictive power of the model. Intriguingly, signalling networks with similar feedback terms can potentially exhibit damped and even sustained oscillations depending on network topology and parameter values \cite{Kholodenko2010}.
	
	\subsection{Flux of active YAP}
	
	Next, we describe the spatiotemporal evolution of YAP that is controlled through two Rho-dependent pathways: nuclear localization driven by the actin cytoskeleton and myosin-contractility-regulated inactivation through phosphorylation, see e.g. \cite{Das2016}. Both pathways are effectively controlled by Rho, which regulates the polymerization of actin via mDia and myosin activity via Rho-associated protein kinase (ROCK) \cite{Sun2016,Burridge2019,Narumiya2009}. \par 
	
	First, we consider the flux of active YAP to the nucleus $\emph{\textbf{J}}_{Y_\mathrm{act}}^\mathrm{nuclear \ tension}$ due to the mechanical tension generated by the cytoskeleton. Based on recent experimental data that suggests deformation of the nucleus is key to activating YAP \cite{Elosegui-Artola2017, Aureille2019, Shiu2018}, we assume that activation of Rho generates tension that stretches the cell nucleus and drives the nuclear localization of active YAP. Such forces are known to originate from perinuclear stress fibers \cite{Shiu2018}, and stress fiber formation is thought to be controlled by Rho \cite{Narumiya2009}. Our assumption of YAP-specific nuclear transport is supported by experimental data from Kofler et al.~\cite{Kofler2018}. Their observations of nuclear transport specific to TAZ led them to hypothesize that force-induced changes in the nuclear pore complex may `stimulate sequence-specific, mediated import of TAZ'~\cite{Kofler2018}.
		
	We therefore relate the force active YAP is experiencing, $\emph{\textbf{F}}$, to its flux towards the nucleus,
	
	\begin{equation}\label{eq6}
	\emph{\textbf{J}}_{Y_\mathrm{act}}^\mathrm{nuclear \ tension} =  
	Y_\mathrm{act}\emph{\textbf{v}}_{Y_\mathrm{act}} = Y_\mathrm{act}\frac{\emph{\textbf{F}}}{\xi},
	\end{equation}	
	where we assume overdamped dynamics, which means that the velocity $\emph{\textbf{v}}_{Y_\mathrm{act}}$ is simply related to $\emph{\textbf{F}}$ through the friction coefficient $\xi$. $\emph{\textbf{F}}$ is an effective measure of the amount of nuclear tension and thus implicitly accounts for cell deformation, stretching and compression.
	Since the magnitude of the mechanical stress $|\sigma|$ associated with the tension is normally expressed to be proportional to the concentration of actomyosin that is controlled by the active form of Rho, $|\sigma| \sim A_{\mathrm{act}}$~\cite{tao2015active,kopfer2020mechanochemical,husain2017emergent,banerjee2017actomyosin,streichan2018global,munster2019attachment,gross2019guiding}, we expect the force to be proportional to the gradient of active Rho. 
	
	We assume that the tension in the stress fibers is controlled by the value of this gradient averaged over the whole cell and obtain the expression $|\emph{\textbf{F}}| \sim |\grad \cdot \sigma| = \langle \abs{\grad{A_\mathrm{act}}}\rangle$, with spatial averaging defined via $\langle X \rangle = \int_{S_\mathrm{cell}}X\differential S/ \int_{S_\mathrm{cell}}\differential S = \left(\int_{S_\mathrm{cell}}X\differential S\right)/S_\mathrm{cell}$; depending on the dimensionality of the model, $S_\mathrm{cell}$ is the length, area or volume of the cell. This form of $|\emph{\textbf{F}}|$ is consistent with the accepted paradigm that as the cell polarizes and the gradients in Rho are established, tension is enhanced within the actin cable, leading to cell stretching and spreading on the substrate~\cite{banerjee2017actomyosin,streichan2018global,munster2019attachment,gross2019guiding}; note that here we assume linear proportionality between the concentrations of active Rho  and actin.
	\subsubsection*{Coupling of active YAP to Rho through the effective charge $q$}
	We assume that $\emph{\textbf{F}}$ drives active YAP to the nucleus and describe this through the attractive harmonic potential $\phi$: $\emph{\textbf{F}} = - \langle \abs{\grad{A_\mathrm{act}}}\rangle \grad{\phi}$. With this in mind, we can rewrite eq. \eqref{eq6} for the flux as $\emph{\textbf{J}}_{Y_\mathrm{act}}^\mathrm{nuclear \ tension} = -Y_\mathrm{act}\xi^{-1}\left\langle \abs{\grad{A_\mathrm{act}}}\right\rangle\grad{\phi}$. This equation is equivalent in form to that for the flux of charged particles in an electrical potential, see e.g., \cite{Levich1962}, and the factor $\xi^{-1}\left\langle \abs{\grad{A_\mathrm{act}}}\right\rangle$ can be interpreted as the mobility of active YAP in the field due to $\phi$; following the analogy, we can think of $q = \xi^{-1}$ as an effective charge; if all quantities were dimensional, $q$ would have units of $\SI{}{m^2.s^{-1}}C_{0 \mathrm{A}}^{-1}$.
	
	For simplicity, here we consider the cell as a one-dimensional system and use a harmonic potential $\phi$ whose gradient is
	\begin{equation}\label{eq8}		
	\frac{\differential \phi}{\differential x} = 
	\begin{cases}
	2\frac{x - \frac{x_\mathrm{nucl \ front}+x_\mathrm{nucl \ back}}{2}}{x_\mathrm{nucl \ back}  - x_\mathrm{nucl \ front}}
	& \text{if} \ t > 0 \text{ and } x_\mathrm{nucl \ front} \leq x \leq x_\mathrm{nucl \ back}\\
	0
	& \text{otherwise.}
	\\
	\end{cases}   	  			  	  
	\end{equation}
	$x$ is a spatial coordinate nondimensionalized by the cell length, $L = \SI{1e-5}{m}$; $x_\mathrm{nucl \ i}$ are the coordinates of the front and the back of the nucleus; here, we fix them to $x_{\mathrm{nucl \ front }} = 1/3$ and $x_{\mathrm{nucl \ back }} = 2/3$. With this choice of $\phi$, its gradient is nondimensional and has a maximum absolute value that is normalized to unity.
	
	\subsubsection*{Flux of active YAP in the absence of a stimulus}
	
	Experimental data indicates that when cells are subjected to conditions that do not stimulate division and motility, i.e., when their adhesive area is low and/or they are seeded on a soft substrate, they exhibit a ratio of nuclear to cytoplasmic YAP, $R = \int_{S_\mathrm{nucl}} \left(Y_\mathrm{act} + Y_\mathrm{inact} \right)\differential S/\left[1 - \int_{S_\mathrm{nucl}} \left(Y_\mathrm{act} + Y_\mathrm{inact} \right)\differential S \right]$, approximately equal to $1$, see e.g. \cite{Aureille2019, Das2016, Peyret2019}. Inactive YAP is sequestered to the cytoplasm where it binds to the protein 14-3-3 \cite{Yu2015, Dobrokhotov2018}, which means that if the nucleus is smaller in volume than the cytosol, the nuclear-to-cytoplasmic YAP ratio can only be $\approx 1$ if active YAP is attracted to the nucleus even in the absence of Rho activation. This suggests an additional term in the flux of $Y_\mathrm{active}$,
	\begin{equation}\label{eq9}
	\emph{\textbf{J}}_{Y_\mathrm{act}}^{\mathrm{rest}}=	-q_0Y_\mathrm{act}\grad{\phi},			
	\end{equation}
	where the effective charge in the absence of stimulation ($q_0$) would be measured in units of $\SI{}{m.s^{-1}}$ if all quantities were dimensional.
	
	The total flux of active YAP is the sum of the contributions in eqs. \eqref{eq6} and \eqref{eq9} and the diffusional flux of the species,
	\begin{equation}\label{eq10}
	\emph{\textbf{J}}_{Y_\mathrm{act}} = \emph{\textbf{J}}_{Y_\mathrm{act}}^\mathrm{nuclear \ tension} +\emph{\textbf{J}}_{Y_\mathrm{act}}^{\mathrm{rest}} - D_{Y_\mathrm{inact}}\grad{Y_\mathrm{inact}} = -\left(q_0 + q\left\langle \abs{\grad{A_\mathrm{act}}}\right\rangle\right)Y_\mathrm{act}\grad{\phi} -D_{Y_\mathrm{act}}\grad{Y_\mathrm{act}}.			
	\end{equation}
	
	\subsection{Flux of inactive YAP}
	
	Similarly, the flux of inactive YAP contains a diffusional contribution and one due to its sequestering by 14-3-3, which we model via a term that repels it from the nucleus. Since YAP phosphorylated at serine 112 forms a complex with 14-3-3 \cite{Das2016}, we assume that this interaction is unaffected by mechanical stimuli and introduce a term that repels inactive YAP from the nucleus regardless of the polarization of the cell with respect to active Rho,
	\begin{equation}\label{eq11}
	\emph{\textbf{J}}_{Y_\mathrm{inact}} = 	q_1 Y_\mathrm{inact} \grad{\phi} - D_{Y_\mathrm{inact}}\grad{Y_\mathrm{inact}},
	\end{equation}
	where the effective charge for inactive YAP ($q_1$) has units of $\SI{}{m.s^{-1}}$.
	
	Since activated YAP is transferred to the nucleus where it binds to the transcription factor TEAD \cite{Elosegui-Artola2017}, its diffusion in this region is impaired. An approximate way of taking that into account is to assume that $D_{Y_\mathrm{act}}$ varies smoothly from its cytoplasmic value to that for the nuclear region, and in particular that it is given a hyperbolic tangent (see the SI for the precise form we use). Note that this assumption likely overestimates the effect of YAP-TEAD binding as the 1D model only allows diffusion through the nucleus but not around it. For the same reason, we fix the diffusion coefficient of the inactive form of YAP to the value measured for it in the cytoplasm as an average for the two species, $D_{Y_\mathrm{cyto}}$ \cite{Singh2017}. \par
	
	\subsection{Rate of YAP generation}
	
	Finally, the rate of generation of active YAP, $f_Y$, includes two contributions - one due to first-order self-activation, which is proportional to the concentration of inactive YAP, and one due to deactivation via, e.g., the LATS pathway, see \cite{Halder2012, Dobrokhotov2018, Moya2019}. Experimental data indicates that inhibiting myosin contractility either directly with blebbistatin or via inhibiting ROCK with Y-27632 causes an increase in the relative abundance of the inactivated form of YAP that is phosphorylated at serine 112 \cite{Das2016}. For this reason, we introduce a term in the rate of interconversion between active and inactive YAP that favours the active form at high myosin contractilities, and vice versa. Myosin is indirectly activated by Rho, and by the same argument we applied to modelling Rho-induced nuclear tension, we assume that myosin contractility is proportional to the average gradient in active Rho. Furthermore, we assume that the myosin-regulated YAP activation saturates, which we model with a first-order Hill function. Combining both contributions, we arrive at the following expression for the rate of chemical generation of active YAP:
	\begin{equation}\label{eq13}
	f_Y(A_\mathrm{act}, Y_\mathrm{act}, Y_\mathrm{inact}) = c_6 Y_\mathrm{inact} - c_7 Y_\mathrm{act}\left(1 - \frac{c_8 \langle \abs{ \grad{A_\mathrm{act}}}\rangle }{c_{9} + c_{10} \langle \abs{\grad{A_\mathrm{act}}}\rangle}\right);
	\end{equation}
	to keep the units consistent, $c_6$ and $c_7$ are measured in $\mathrm{s^{-1}}$, $c_8$ and $c_{10}$ - in $C_{0 \mathrm{A}}^{-1}L$, and $c_9$ is dimensionless. The values of the parameters for the rate of YAP generation are listed in Table~\ref{tbl:additional_parameters} in the Supporting Information. \par
	It is worthy of note that due to the presence of the average active Rho gradient in eqs. \eqref{eq10} and \eqref{eq13}, as well as that of the amount of active YAP in the nucleus in eq.~\eqref{eq5}, our reaction-diffusion equations \eqref{eq1}-\eqref{eq4} are partial integro-differential rather than simply differential equations; moreover, they are non-linear.
	
	\begin{table*}[t]
		\small
		\caption{\ Key parameters based on other sources. Note, however, that we use values of the chemical rate constants $c_1, c_2$ and $c_4$ that are higher than those of Mori et al. by an order of magnitude. We do so because it appears that the simulations in that paper were conducted with a diffusion coefficient for active Rho ten times lower than the value given in the text; slower diffusion of active Rho translates into a more robust gradient in $A_\mathrm{act}$. Increasing the rate constants by the same factor leads to approximately the same sharp `pinned wave' concentration profile as the one in Mori et al.'s Fig. 2a \cite{Mori2008}.}
		\label{tbl:main_parameters}
		\makegapedcells
		\begin{tabular*}{\textwidth}{@{\extracolsep{\fill}}lllll}
			\hline
			\makecell[l]{Parameter} & \makecell[l]{Meaning} & \makecell[l]{Value} & \makecell[l]{Units} & \makecell[l]{Reference} \\
			\hline
			\makecell[l]{$D_\mathrm{A_\mathrm{act}}$}& \makecell[l]{diffusion coefficient of active Rho} & 		    \makecell[l]{$10^{-13} $} & \makecell[l]{$\SI{}{m^2.s^{-1}} $} & \makecell[l]{\cite{Postma2004}} \\
			\hline
			\makecell[l]{$D_\mathrm{A_\mathrm{inact}}$} & \makecell[l]{diffusion coefficient of inactive Rho} & 		\makecell[l]{$10^{-11} $} & \makecell[l]{$\SI{}{m^2.s^{-1}} $} & \makecell[l]{\cite{Postma2004}} \\					
			\hline
			\makecell[l]{$D_\mathrm{Y_\mathrm{cyto}}$} & \makecell[l]{diffusion coefficient of YAP in the cytoplasm} & 		\makecell[l]{$8 \times 10^{-11}$} & \makecell[l]{$\SI{}{m^2.s^{-1}} $} & \makecell[l]{\cite{Singh2017}} \\
			\hline
			\makecell[l]{$D_\mathrm{Y_\mathrm{nucl}}$} & \makecell[l]{diffusion coefficient of YAP in the nucleus} & 		\makecell[l]{$4.5 \times 10^{-12}$} & \makecell[l]{$\SI{}{m^2.s^{-1}} $} & \makecell[l]{\cite{Singh2017}} \\
			\hline
			\makecell[l]{$L$} & \makecell[l]{cell length} & \makecell[l]{$10^{-5}$} & \makecell[l]{$\SI{}{m} $} & \makecell[l]{\cite{Mori2008}} \\
			\hline
			\makecell[l]{$c_1$} & \makecell[l]{base activation rate for Rho} & \makecell[l]{$0.67$} & \makecell[l]{$\SI{}{s^{-1}} $} & \makecell[l]{\cite{Mori2008}} \\
			\hline
			\makecell[l]{$c_2$} & \makecell[l]{Hill function parameter for Rho} & \makecell[l]{10} & \makecell[l]{$\SI{}{s^{-1}} C_{0 \mathrm{A}}^{-2}$} & \makecell[l]{\cite{Mori2008}} \\
			\hline
			\makecell[l]{$c_3$} & \makecell[l]{Hill function parameter for Rho} & \makecell[l]{10} & \makecell[l]{$C_{0 \mathrm{A}}$} & \makecell[l]{\cite{Mori2008}} \\
			\hline
			\makecell[l]{$c_4$} & \makecell[l]{base deactivation rate for Rho} & \makecell[l]{10} & \makecell[l]{$\SI{}{s^{-1}} $} & \makecell[l]{\cite{Mori2008}} \\
			\hline													
		\end{tabular*}
		\nomakegapedcells
	\end{table*}
	
	\section{Methods}			
	
	We solve equations \eqref{eq1}-\eqref{eq4} by performing updates of the integral terms at discrete intervals $\Delta t$ and treating them as constant within each such interval. This allows us to use the 1D partial differential equation solver built into MATLAB R2020a (9.8.0.1451342), pdepe, which employs a second-order finite-element discretization in space and a variable-step, variable-order algorithm for time integration (ode15s), see Ref.~\cite{Skeel1990}; pdepe ensures flux continuity in the solution region~\cite{pdepe2020}. We use a uniform grid of 250 mesh points for spatial discretization and set $\Delta t$ such that it is considerably smaller than the time scale for the fastest process that occurs in a system with a particular set of parameters; $\Delta t$ thus varies in the range $\SI{1e-4}{s}-\SI{1e-1}{s}$.
	We assume that the total amounts of YAP and Rho in the cell are conserved and therefore impose no-flux boundary conditions at both ends of the cell for all species,
	\begin{equation}\label{eq14}
	\textbf{n}\cdot \grad{Y_\mathrm{i}}|_{x = 0, x = L} =  \textbf{n}\cdot \grad{A_\mathrm{i}}|_{x = 0, x = L} = 0,
	\end{equation}
	where $\textbf{n}$ is the outward normal to the cell boundaries.
	
	\subsection{Equilibration and initial perturbations}
	
	We investigate the dynamics of YAP localization upon the introduction of a perturbation in the concentration of active Rho. To do this, we first let the system equilibrate in the absence of any stimuli. We use homogeneous initial conditions for the GTPase: $A_\mathrm{inact} = 20$ and $A_\mathrm{act} = A_\mathrm{act}^- = 2.683$, $A_\mathrm{act}^-$ being the lower stable steady concentration that corresponds to this value of $A_\mathrm{inact}$, obtained as a solution to the equation $	f_A (A_\mathrm{act}, A_\mathrm{inact}, Y_\mathrm{act})|_{c_5 = 0}$. The simulation starts with a homogeneous YAP distribution that includes no active form, $Y_\mathrm{act} = 0$ and $Y_\mathrm{inact} = 1$. \par 
	
	Due to the terms proportional to $q_0$ and $q_1$, the YAP concentrations evolve so that the active form has a peak in the nucleus and the inactive form is depleted from it. The concentration profiles of the two forms of Rho remain homogeneous but change because, due to the $c_5$ term, $A_\mathrm{act}^-$ is no longer a stationary concentration. All concentration distributions reach a steady state within $\sim \SI{10}{s}$. \par				
	
	We simulate the equilibration of the system for each set of kinetic parameters (see Figure~\ref{fgr:C_snapshots_equilibration_q=4_2_YAP_ratio_t_sweep}(c) in the SI). We then apply a localized parabolic initial perturbation to the stationary Rho concentration profiles obtained from this simulation, $A_\mathrm{i \ stat}$. The form of the perturbation preserves the overall amount of Rho GTPase in the system and obeys no-flux boundary conditions,
	\begin{equation}\label{eq15}
	\delta A_\mathrm{act} = 
	A_\mathrm{pert}A_\mathrm{act \ stat} \left(1 - (x/L_\mathrm{pert})^2\right)\mathrm{Heaviside}(L_\mathrm{pert} - x)
	\end{equation}
	with
	\begin{equation}\label{eq16}
	A_\mathrm{act}|_{t=0} =  A_\mathrm{act \ stat} + \delta A_\mathrm{act}
	\end{equation}
	and 
	\begin{equation}\label{eq17}
	A_\mathrm{inact}|_{t=0} =  A_\mathrm{inact \ stat} - \delta A_\mathrm{act}.
	\end{equation}	
	
	Unless otherwise noted, we use $A_\mathrm{pert} = 0.75$ and $L_\mathrm{pert} = 0.15$; the effect of varying $A_\mathrm{pert}$ is shown in Figure~\ref{fgr:R_t_diff_A_pert_trans_stim} in the SI. In most simulations, localized perturbations of this form with a sufficiently large amplitude generate a polarized profile of $A_\mathrm{act}$ with a region of high concentration $A_\mathrm{act}$ in the vicinity of the initial stimulus, a region of low $A_\mathrm{act}$ on the other end of the cell, and a steep gradient in between (`pinned wave'). This happens because, as per eq.~\eqref{eq5}, the production of active Rho is an autocatalytic reaction, and a local increase of $A_\mathrm{act}$ sufficiently large not to be dispersed by diffusion leads to its accumulation. At the same time, the positive feedback of $A_\mathrm{act}$ on itself is limited because of the second-order Hill function in eq. \eqref{eq5}, which leads to a plateau in $A_\mathrm{act}$. This simulates the behaviour of a cell which receives an external stimulus and acquires polarization due to spatial variations in the concentration of GTP-bound Rho.\par
	
	We examine the effect of the form of the initial conditions on the system's dynamics by using an alternative way of inducing a gradient in Rho, following Mori et al.~\cite{Mori2008}. At short times, we add a transient local stimulus term of the form $k_\mathrm{stim}A_\mathrm{inact}$ to $	f_A (A_\mathrm{act}, A_\mathrm{inact}, Y_\mathrm{act})$
	\begin{equation}\label{eq18}
	k_\mathrm{stim} = s(t)\left(1 + \cos(\uppi x)\right)\mathrm{Heaviside}(L/10 - x),
	\end{equation} 
	where
	\begin{equation}\label{eq19}
	s(t) = 
	\begin{cases}
	S_\mathrm{Ampl}/2
	& \text{if } 0 \leq t \leq t_1; \\
	S_\mathrm{Ampl}/4\left[1 + \cos\left(\uppi \frac{t-t_1}{t_2 - t_1}\right)\right]
	& \text{if } t_1 \leq t \leq t_2.\\
	0
	& \text{otherwise,}
	\\
	\end{cases}
	\end{equation} 
	with $S_\mathrm{Ampl} = 0.5, t_1 = 20$ and $t_2 = 25$ \cite{Mori2008}; as with $c_1, c_2$ and $c_4$, we increase the value of $S_\mathrm{Ampl}$ tenfold from the one reported in the paper to compensate for the mistake in the diffusion coefficient of active Rho in Ref. \cite{Mori2008}. The two types of initial conditions - eqs. \ref{eq15}-\ref{eq17} vs. eqs. \ref{eq18}-\ref{eq19} are compared in Figure~\ref{fgr:R_t_diff_A_pert_trans_stim} in the SI.
	
	\section{Results}
	
	Here we discuss the specific predictions of the model for different parametrizations of the coupling between YAP and Rho: (1) non-oscillatory and (2) oscillatory YAP/Rho dynamics at mechanical stimuli ($q$) of fixed magnitude; (3) YAP/Rho dynamics for time-dependent $q$.			
	
	\subsection{Non-oscillatory YAP/Rho dynamics at a fixed mechanical stimulus magnitude ($q$)}
	
	Experimental studies show that the ratio of nuclear-to-cytoplasmic YAP ($R$) is close to unity for unstimulated cells and can increases up to severalfold upon the application of a mechanical stimulus that most typically increases the cell's adhesive area - see e.g. Ref. \cite{Peyret2019}, where the ratio varies from $\sim 1$ to $\approx 3$, and Ref. \cite{Nardone2017}, where it saturates at $\approx 5$. \par
	
\begin{figure}
		\centering					
		\subfloat[]
		{
			\includegraphics[width=8cm]{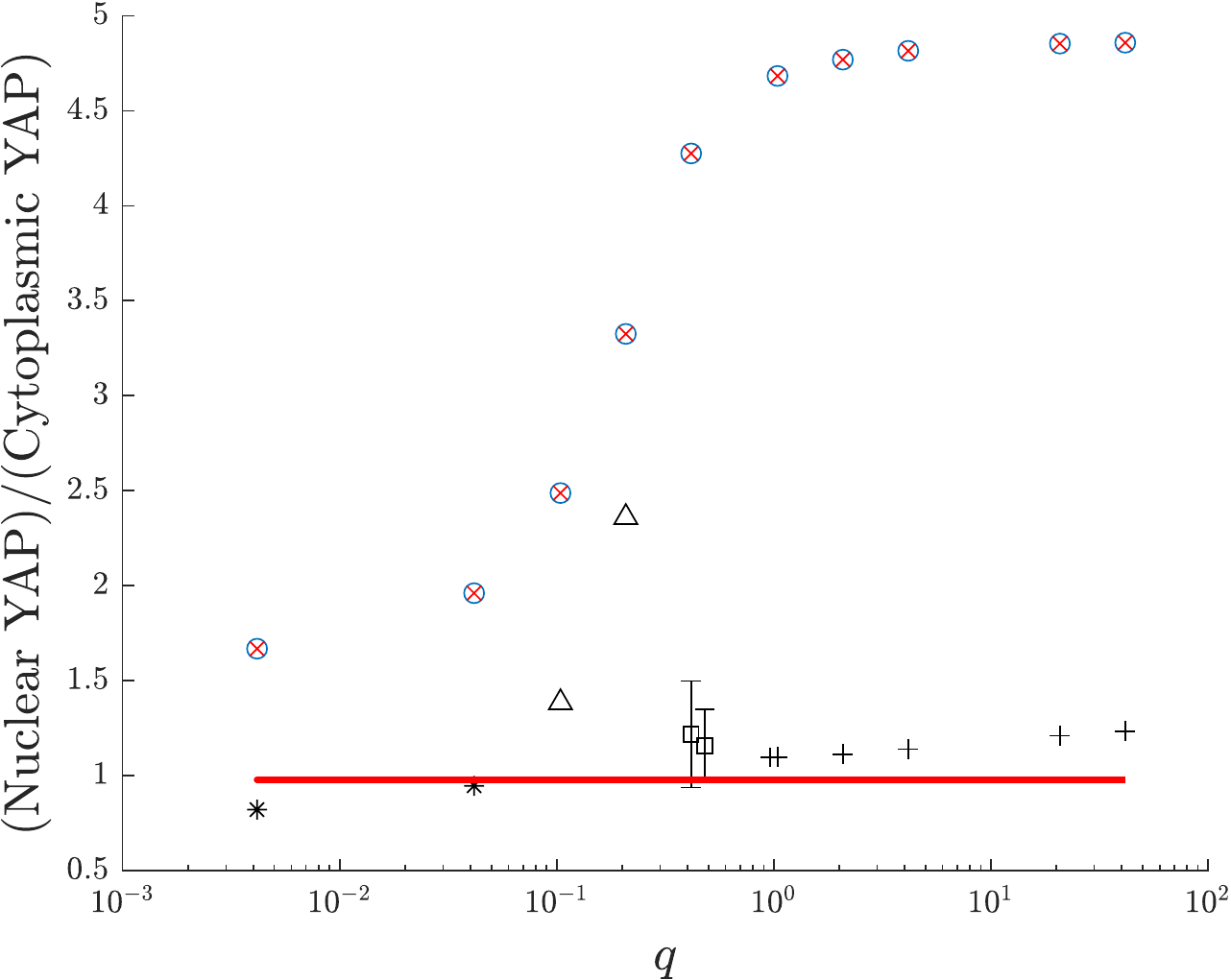}				
		}
		\\
		\subfloat[]
		{
			\includegraphics[width=8cm]{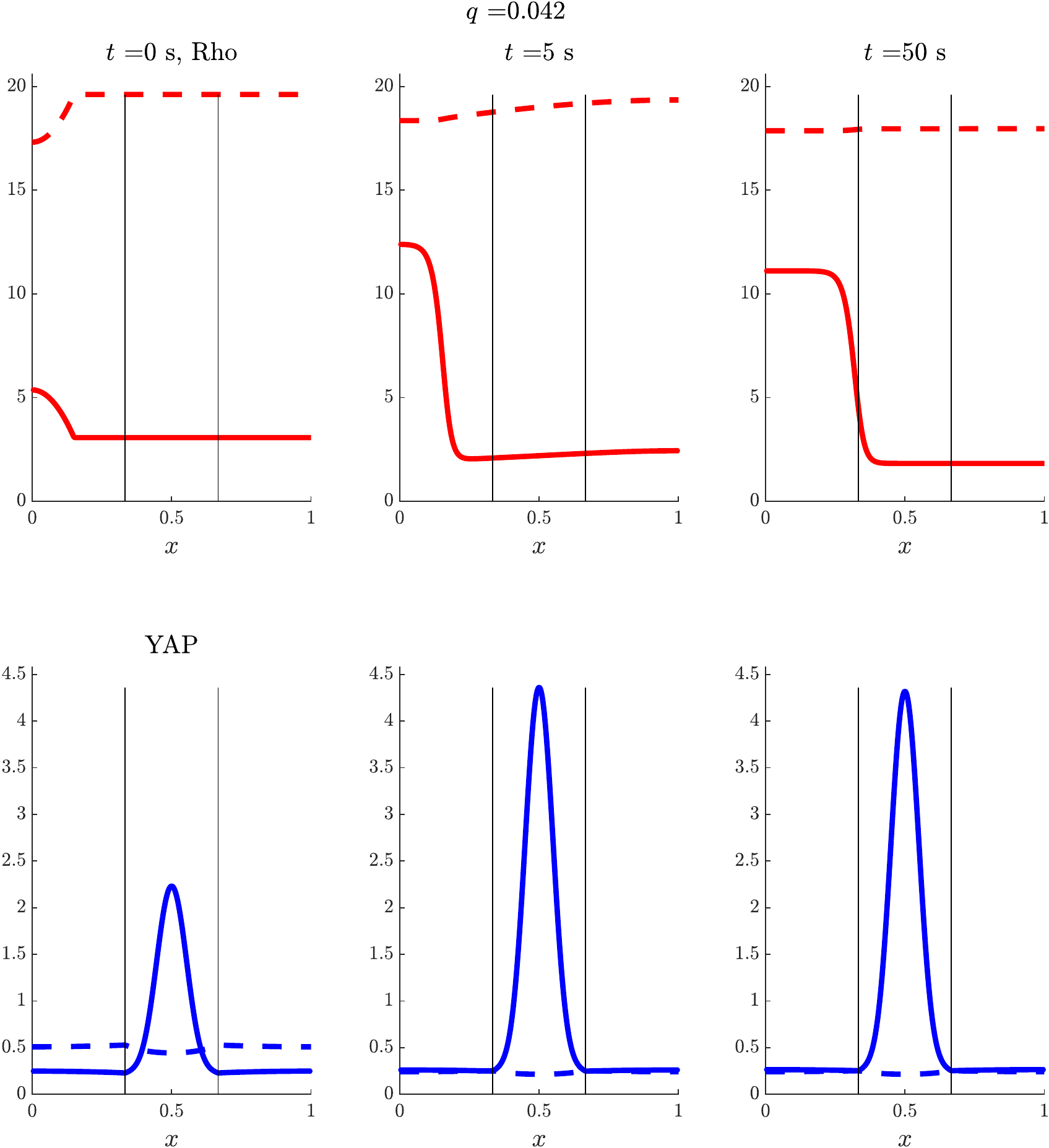}
		}	
		\caption
		{
			\textbf{a}. Stationary nuclear-to-cytoplasmic YAP ratios ($R$) for simulations at different intensities of the mechanical stimulus $q$. The red line gives $R$ for the case with no stimulus in which the concentration profiles of both forms of Rho remain homogeneous; the blue circles are obtained through simulations with a parabolic initial perturbation; the red crosses represent the YAP ratios in in simulations with a transient initial stimulus as per eqs.~\eqref{eq18}-\eqref{eq19}. The YAP-Rho coupling for all these data points is weak and the parameters are: $c_5 = $1, $c_6 = $1, $c_7 = $1, $c_8 = $3.5, $c_9 = $1, $c_{10} = $5, $q_0 = $2.5, $q_1 = $2, $Y_\mathrm{act \ nucl \ threshold}$ =1. The asterisks represent simulations in which the final concentration profile exhibits a nonphysical second region of high $A_\mathrm{act}$ in the nuclear region; the triangles - to cells polarized as in the plot for $t = \SI{50}{s}$ in \textbf{b},  squares - to simulations in which $R(t)$ oscillates in response to sufficiently large perturbations to $A_\mathrm{act}$ (see Section 4.3 below), with error bars indicating the minimum and maximum value $R$ attains in the regime with sustained oscillations, and the pluses stand for simulations that yield damped oscillations in $R(t)$. The parameters for all these data points in black are: $c_5 = $30, $c_6 = $1, $c_7 = $1, $c_8 = $2.5, $c_9 = $1, $c_{10} = $5, $q_0 = $1.25, $q_1$ = 0.5, $Y_\mathrm{act \ nucl \ threshold}$ = 0.4. \textbf{b}. Snapshots of the evolution of the concentrations of the different species in the system at a low value of the mechanical forcing term $q$ and a weak YAP-Rho coupling. Rho (top row) and YAP (bottom row) profiles, with solid lines depicting active forms and dashed lines - inactive forms. The vertical lines indicate the boundaries of the nucleus. As seen in the plot for $t = \SI{0}{s}$, the initial perturbation to the homogeneous Rho concentrations is parabolic.
		}
		\label{fgr:YAP_ratio_q_sweep_C_snapshots_q=0_042}
	\end{figure}

	With this in mind, we choose parameters for our simulations that yield a YAP ratio of $\approx 1$ when the Rho concentration profile in the cell is not polarized: $c_5 = 1, c_6 = 1, c_7 = 1, c_8 = 3.5, c_9 = 1, c_{10} = 5, q_0 = 2.5, q_1 = 2$, see the red line in Figure~\ref{fgr:YAP_ratio_q_sweep_C_snapshots_q=0_042}. We begin by considering strictly positive feedback of YAP on Rho, $Y_\mathrm{act \ nucl \ threshold} = 1$. Upon the application of a sufficiently strong perturbation, the system forms a stationary polarized state and the gradient in Rho increases the amount of active YAP in the nucleus both by directly increasing the effective depth of the potential well (eq.~\eqref{eq6}) and by favouring the formation of the active form through the Hill-function term in eq.~\eqref{eq13}. \par 
	
	We perform a set of simulations in which we vary the value of $q$, and in this way explore the dynamics of YAP and Rho relaxation upon application of a constant mechanical stimulus. We then calculate the stationary nuclear-to-cytoplasmic YAP ratios $R$ at $t = \SI{250}{s}$, when the system has reached a steady state (see Figure~\ref{fgr:YAP_ratio_q_sweep_C_snapshots_q=0_042}). As the figure indicates, $R$ increases steeply in the range $q \sim 10^{-2}-1$ and plateaus at higher $q$. Moreover, the stationary YAP ratios are the same regardless of whether we employ a parabolic initial perturbation in concentration (blue circles) or a transient initial stimulus (red crosses), although the system takes a different path to the stationary state.
		
	\subsection{Oscillations in YAP and Rho activity at fixed $q$}
	
\begin{figure}
		\centering					
		\subfloat[]
		{
			\includegraphics[width=11cm]{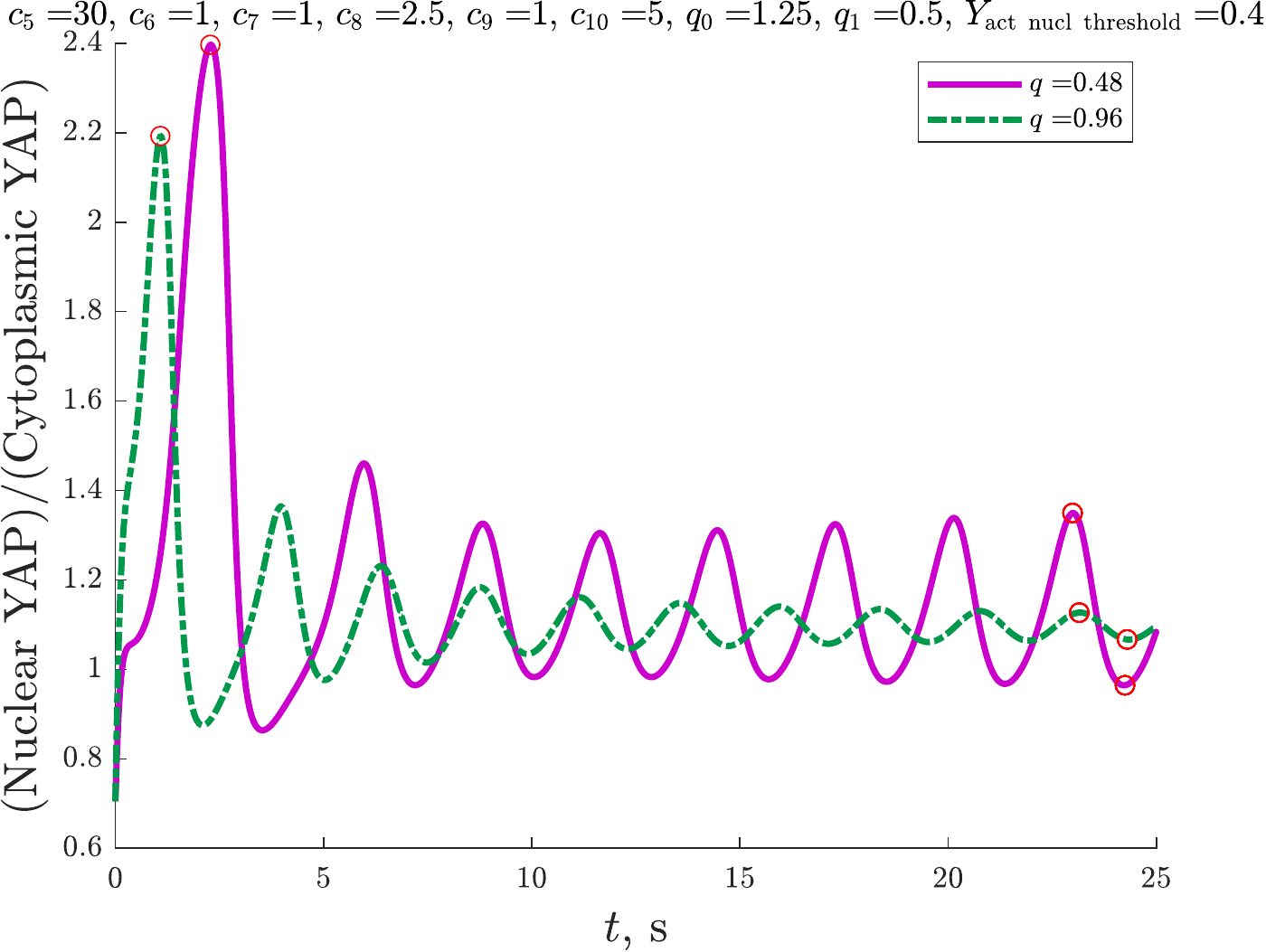}				
		}
		\\
		\subfloat[]
		{
			\includegraphics[width=8cm]{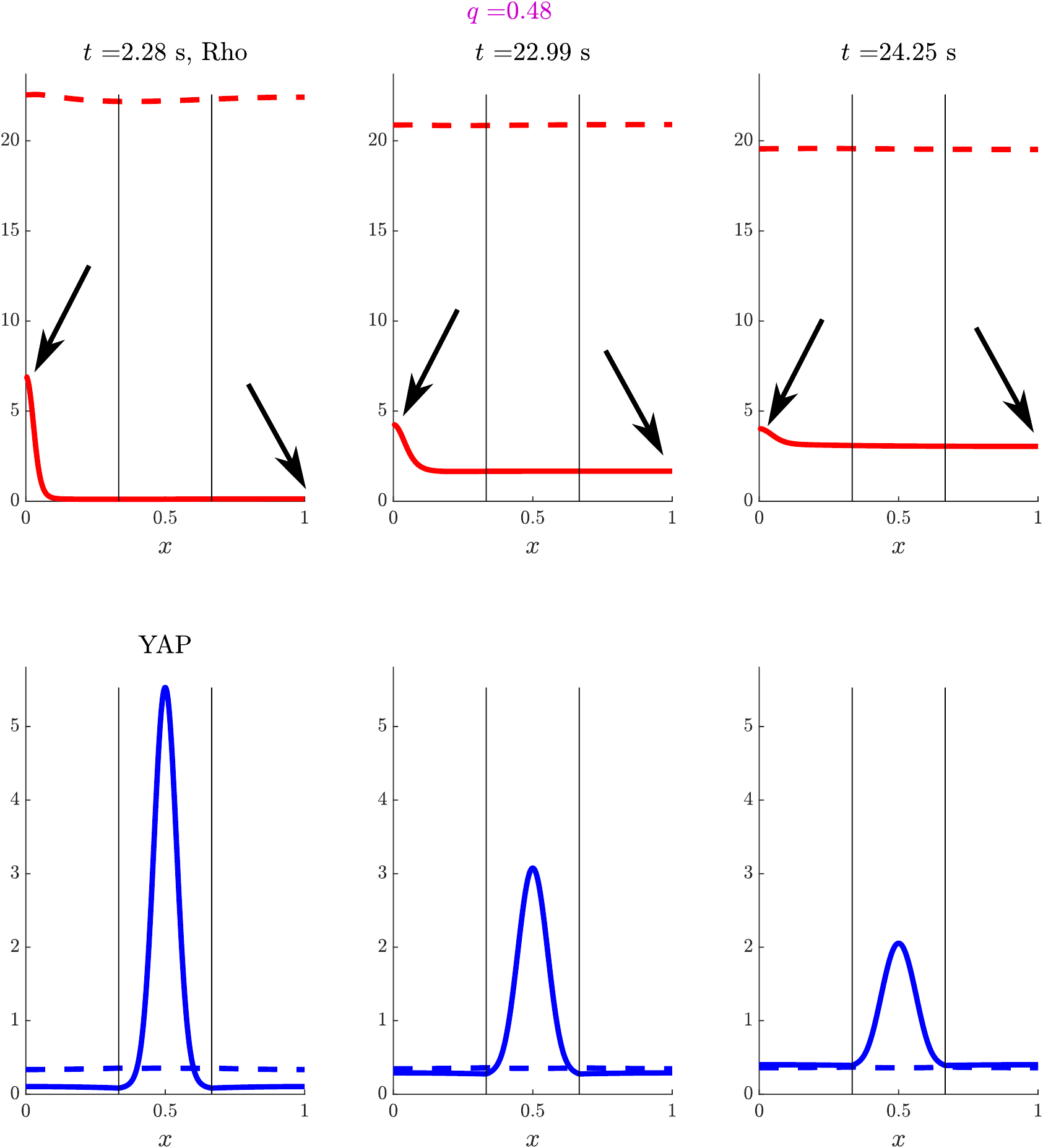}				
		}						
		\subfloat[]
		{
			\includegraphics[width=8cm]{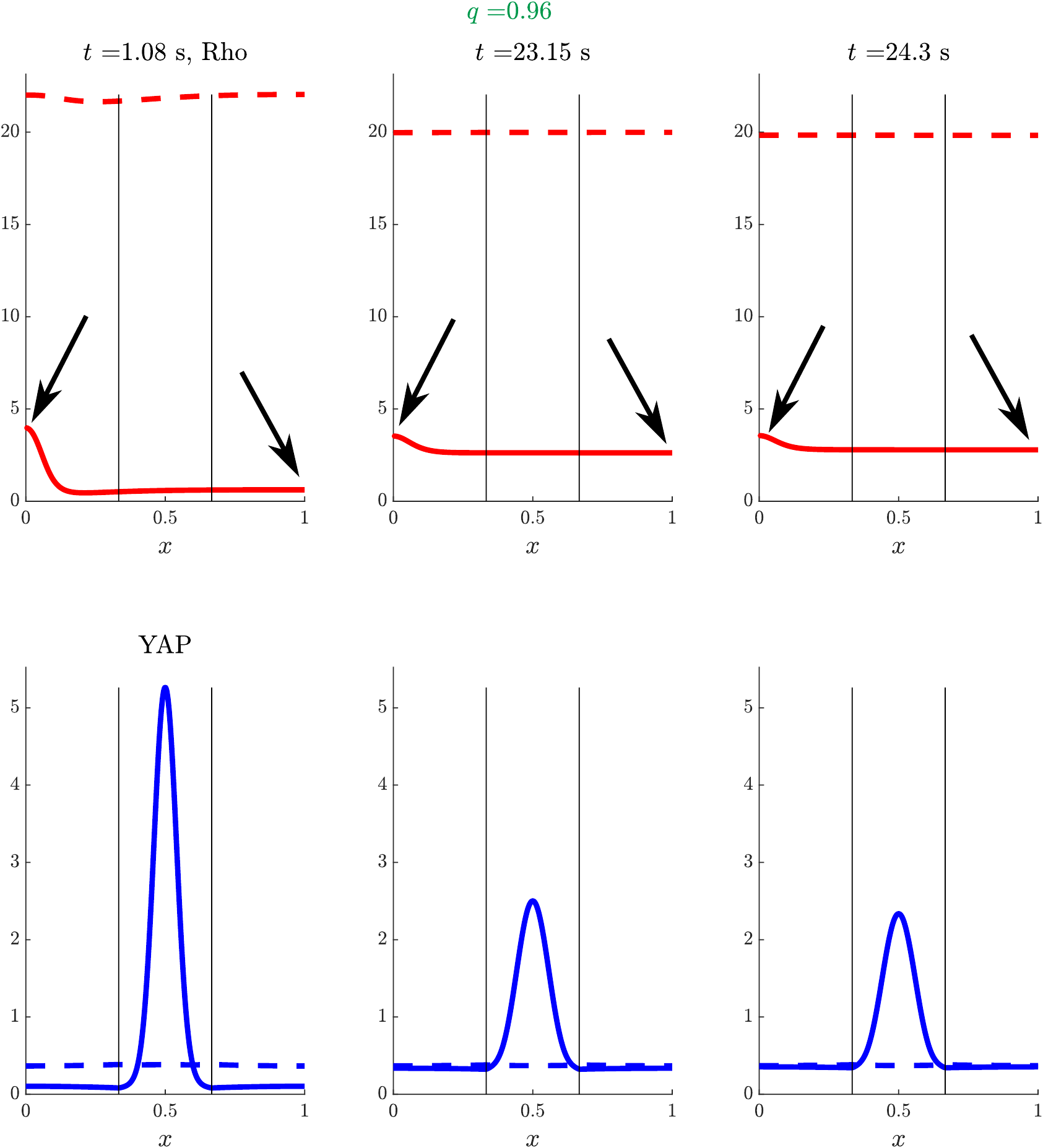}
		}
		\caption
		{
			Results from simulations with strong feedback of YAP on Rho ($c_5 = 30$) that switches sign at $Y_{\mathrm{act \ nucl \ threshold}} = 0.4$.
			\textbf{a}. Nuclear-to-cytoplasmic YAP ratios for $q = 0.48$ (solid line) and $q = 0.96$ (dotted line). Note that the lower $q$ value leads to sustained oscillations in the ratio, whereas the higher one - to damped oscillations.
			\textbf{b}-\textbf{c} As in the rest of the text, Rho profiles are in the top row and YAP profiles - in the bottom one, with solid lines depicting acting forms and dashed lines - inactive forms. The vertical lines indicate the boundaries of the nucleus.
			\textbf{b-c}. Snapshots of the concentration distributions for the simulation at $q = 0.48$ (\textbf{b}) and $q = 0.96$ (\textbf{c}) illustrated in \textbf{a}, taken at some of the peaks in $R$.
			The damping of $R$ in \textbf{c} comes from the damping in $\langle \abs{\grad{A_\mathrm{act}}}\rangle$ (controlled by the values of $A_\mathrm{act}$ at the endpoints, as indicated by the arrows in the plot), which is in turn caused by the stronger feedback between YAP and Rho at higher $q$, see eq.~\eqref{eq6}. Red circles mark the points in time at which the concentration profile snapshots are taken.
		}
		\label{fgr:sust_damped_osc}
	\end{figure}

	Next, we study the nuclear-to-cytoplasmic YAP ratio for systems with both positive and negative feedback of YAP on Rho. This is motivated by recent experiments showing the possibility of the two feedbacks mediated by ARGHGAP proteins~\cite{Mason2019}. Interestingly, by accounting for these feedbacks the system shows both damped and sustained oscillations of the YAP nuclear-to-cytoplasmic ratio for a range of the force magnitude $q$ and the strength of coupling between YAP and Rho. Oscillations can arise in our model as follows: for $Y_{\mathrm{act \ nucl \ threshold}} < 1$, feedback of nuclear YAP on Rho changes sign if $Y_{\mathrm{act \ nucl}} > Y_{\mathrm{act \ nucl \ threshold}}$. This reduces the gradient in active Rho, causing $Y_{\mathrm{act \ nucl}}$ and the nuclear-to-cytoplasmic ratio $R$ to go down because of the weaker driving force for YAP nuclear translocation (eq.~$\eqref{eq6}$) and YAP activation (eq.~$\eqref{eq13}$). For suitable parameter values, polarization is not fully lost in this process ($\langle \abs{ \grad{A_\mathrm{act}}}\rangle \neq 0 $) such that once $Y_{\mathrm{act \ nucl}} < Y_{\mathrm{act \ nucl \ threshold}}$, YAP activates Rho again, causing the gradient in active Rho to increase and $R$ to follow suit until $Y_{\mathrm{act \ nucl}}$ reaches $Y_{\mathrm{act \ nucl \ threshold}}$ and the cycle is reset. We illustrate this process through snapshots of the concentration profiles for two simulations in Figure~\ref{fgr:sust_damped_osc}: one of these (for $q = 0.48$) results in sustained oscillations, whereas the other (for $q = 0.96$) yields damped oscillations. \par
	
	Apart from $Y_{\mathrm{act \ nucl \ threshold}}$, the system's behaviour is sensitive to the coupling constants that govern the effect of YAP on Rho ($c_5$) and that of Rho on YAP ($q$). We have found oscillations for $c_5 = 30$, a value much greater than the simulations in Figure~\ref{fgr:YAP_ratio_q_sweep_C_snapshots_q=0_042} that exhibit non-oscillatory dynamics at $c_5 = 1$. For oscillations to occur, $c_5$ needs to be large enough that $\langle \abs{ \grad{A_\mathrm{act}}}\rangle$ is significantly perturbed, but not so large that nuclear active YAP levels above $Y_{\mathrm{act \ nucl \ threshold}}$ cause active Rho to adopt a uniform profile. \par 
	
	Moreover, the stationary $R$ tends to reach saturation with $q$ (see Figure \ref{fgr:YAP_ratio_q_sweep_C_snapshots_q=0_042}(a)), which means that at high $q$, changes in the term $q \langle \abs{ \grad{A_\mathrm{act}}}\rangle$ in the flux of nuclear YAP, eq. \eqref{eq6}, that fully preserve polarization and, as such, weakly affect $\langle \abs{ \grad{A_\mathrm{act}}}\rangle$, are going to be too weak to cause a significant change in $R$. At the same time, if $q \to 0$, changes in $\langle \abs{ \grad{A_\mathrm{act}}}\rangle$ will be too weak to affect $R$. The oscillatory regime thus requires $q$ to be in the range that roughly corresponds to the region of the curve in Figure~\ref{fgr:YAP_ratio_q_sweep_C_snapshots_q=0_042}(a) with the greatest slope. This is indeed what we see when we perform simulations at different $q$ and fixed other parameters: at $q \to 0$, $R$ reaches a stationary value without oscillating (notably, the profile of Rho develops a second region of high Rho activity), and at $q > 10$ the stationary state is reached within $\sim \SI{10}{s}$ after several damped oscillations. In the intermediate range, $q \sim 1$, illustrated in Figure~\ref{fgr:sust_damped_osc}, we see slowly damped oscillations at $q = 0.96$ and sustained ones at $q = 0.48$, both types with a period of $\sim \SI{1}{s}$. \par 
	
	A pertinent question is whether these oscillations are robust with respect to changing the initial conditions. We tested this by conducting runs for different values of $A_\mathrm{pert}$ in eq.~\eqref{eq15} and using the transient stimuli introduced by Mori et al. \cite{Mori2008}, eq.~\eqref{eq16}-\eqref{eq17}. These tests, which are illustrated in Figure~\ref{fgr:R_t_diff_A_pert_trans_stim} in the Supporting Information, show that perturbations that are too small ($A_\mathrm{pert} \lesssim  0.1$ at $q = 0.48$ and $A_\mathrm{pert} \lesssim  10^{-2}$ at $q=0.96$), do not excite oscillations in $R$. This happens because such perturbations are dispersed by diffusion and do not generate a polarized profile in active Rho; consequently, the concentrations of all species return to the equilibrium values calculated as per Section 4.1. However, if the initial perturbation is strong enough, it generates a pinned wave in $A_\mathrm{act}$ which through its corresponding non-zero $\langle \abs{ \grad{A_\mathrm{act}}}\rangle$ activates YAP. When the level of active nuclear YAP exceeds the threshold value, the system enters an oscillatory state like the ones illustrated in Figure~\ref{fgr:sust_damped_osc}. The comparison between $R(t)$ curves for different initial conditions in Figure~\ref{fgr:R_t_diff_A_pert_trans_stim} shows that the frequency of the oscillations is unaffected by changes in $A_\mathrm{pert}$, and that for the sustained case, the same is true of the amplitude after the initial transient. This is to be expected as due to the form of eq.~\eqref{eq5}, the two attractors in this system are the equilibrated state with homogeneous Rho concentrations and the state with polarized Rho and $Y_{\mathrm{act \ nucl}} = Y_{\mathrm{act \ nucl \ threshold}}$. \par
	
	The oscillations we describe here occur about the value of $R$ at which $Y_{\mathrm{act \ nucl}} = Y_{\mathrm{act \ nucl \ threshold}}$; thus, we can tune the stationary $R$ by changing $Y_{\mathrm{act \ nucl \ threshold}}$. We illustrate this in Figure~\ref{fgr:R_t_q_0_48_Yactnuclthr_0_425} in the Supporting Information, where we change $Y_{\mathrm{act \ nucl \ threshold}}$ from 0.4 to 0.425 in comparison with Figure~\ref{fgr:sust_damped_osc}b and see that in this case, $R(t)$ oscillates about $\approx 1.5$ rather than 1.15.
	
	It is instructive to note that while transient YAP dynamics have been only marginally explored, oscillations in Rho have been reported in several experimental and theoretical studies. Miller and Bement \cite{Miller2009} observed oscillations in Rho activity on a time scale of $\sim \SI{20}{s}$ during cytokinesis in cells in which the GTPase activating protein MgcRacGAP is inactive. They hypothesized that although GAPs typically deactivate Rho proteins, MgcRacGAP in particular also anchors Rho-GTP and thus prevents oscillations from arising. Rho GTPase (RhoA, CDC42, RAC1) activity has also been reported to correlate with cell protrusion formation, and to oscillate on a similar time scale during protrusion-retraction cycles \cite{Machacek2009}. Nikonova et al. \cite{Nikonova2013} formulated an ODE-based model in which Rho activity is controlled by GDP dissociation inhibitors and can exhibit sustained oscillations with a period of $\sim \SI{10}{s}$. \par
	
	More recently, Franklin et al. \cite{Franklin2020} have looked into the dynamic evolution of the YAP nuclear-to-cytoplasmic ratio $R$ and observed large fluctuations on a time scale of hours during monolayer growth. Moreover, sustained oscillations in calcium levels induced via treatment with the drug Thapsigargin correlate with oscillations in nuclear area as well as low-amplitude oscillations in $R$, and the period for all three is of the order of 30 minutes. Additionally, there is indirect evidence that YAP activity may also oscillate in certain contexts. Dequ\'eant et al. \cite{Dequeant2008} reported that Cyr61, a key direct transcriptional target for YAP \cite{Rognoni2019}, is a cyclic gene of the mouse segmentation clock, leading Hubaud et al. \cite{Hubaud2017} to hypothesize that the YAP pathway could be regulated in a periodic fashion. Cyr61 oscillations have a typical time scale of the order of hours \cite{William2007}, and one would expect a similar oscillation period for YAP activity if it is the source of these oscillations. \par				  					  	
	
	\subsection{YAP/Rho dynamics for time-dependent $q$}
	
	Many experimental studies have looked at the effect of cyclic stretching and compression on tissues since such stimuli have relevance, for example, to the study of vascular cells as pulsatile pressure changes in blood pressure lead expansion and compression of large arteries \cite{Hahn2009}. Recently, Landau et al. \cite{Landau2018} have shown that applying cyclic stretching in a culture of endothelial cells and fibroblasts over multiple days leads to translocation of YAP to the nucleus in the fibroblasts and to the alignment of the latter perpendicularly to the direction of stretching. However, we are not aware of any studies that have characterized the effect on YAP activation on the time scale of the applied stretching.
	
\begin{figure}
		\centering					
		\subfloat[]
		{
			\includegraphics[width=11cm]{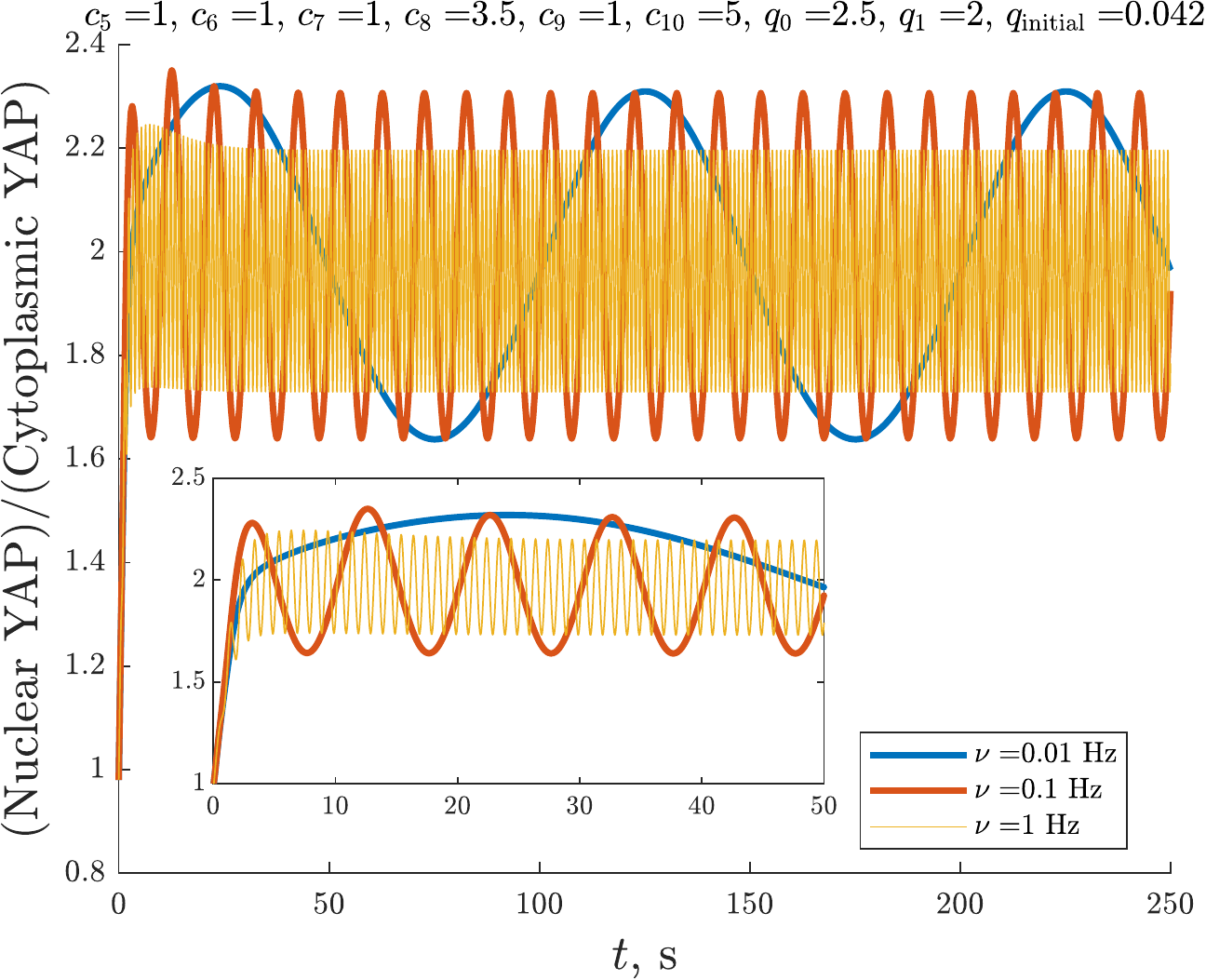}				
		}
		\caption
		{
			Results for $R(t)$ from simulations with $q$ oscillating according to eq.~\eqref{eq22} and other parameters as in Figure~\ref{fgr:YAP_ratio_q_sweep_C_snapshots_q=0_042}b at different frequencies; the inset shows $R(t)$ for short times.
		}
		\label{fgr:N_C_ratio_vs_t_osc_q}
	\end{figure}

	Since our model couples biochemical signalling to mechanics via the effective charge $q$, we can study the effect of time-dependent mechanical stimuli on YAP and Rho activity. In particular, Figure~\ref{fgr:N_C_ratio_vs_t_osc_q} illustrates $R(t)$ for simulations with oscillatory $q(t)$ of the form
	\begin{equation}\label{eq22}
	q(t) = q_\mathrm{initial}\left(1 + \sin(2\uppi\nu t)\right)
	\end{equation}
	at frequencies $\nu$ ranging from $10^{-2}$ to $\SI{1}{Hz}$; the other parameters are the same as in Figure~\ref{fgr:YAP_ratio_q_sweep_C_snapshots_q=0_042}, i.e., they yield $R$ in the same range as experimental data and do not allow for oscillations at a static $q$. \par
	
	The nuclear-to-cytoplasmic YAP ratios for time-dependent $q$ in Figure~\ref{fgr:N_C_ratio_vs_t_osc_q} closely follow the external oscillations implies that our proposed mechanism for YAP activation via Rho-induced nuclear tension could be behind the activation observed in systems subjected to cyclic stretching and compression.					
	
	\section{Discussion and conclusions}
	
	To our knowledge, the model we have presented here is the first one that considers the dynamics of YAP spatial localization in conjunction with its activation through Rho-induced cellular tension. We base our work on a model in which the concentration of active Rho forms a polarized profile in response to localized stimuli \cite{Mori2008}, proposing a two-way coupling between YAP and Rho that implicitly accounts for the influence of mechanical stimuli on YAP activation. We consider several possible types of behaviours of the system that arise in different areas of parameter space, and analyse the model's predictions for the ratio of nuclear to cytoplasmic YAP ($R$), which is used as an experimental measure of YAP activation. \par
	
	At fixed magnitudes of the mechanical stimulus ($q$) and strictly positive feedback of YAP on Rho, we predict that both Rho concentration and $R$ reach a steady value after the initial perturbation to Rho is applied. These stationary values lie on a sigmoidal curve that qualitatively reproduces experimental measurements of $R$ versus cell adhesive area \cite{Nardone2017}. \par
	
	At other values of the kinetic parameters that govern the chemical reactions in the system, in particular if the YAP-Rho coupling is stronger and allows for both positive and negative feedback, our model predicts that both $R$ and the concentrations of the various components exhibit oscillatory behaviour. At both low and high $q$, the system relaxes after a perturbation in Rho via damped oscillations, whereas at intermediate $q$, the same perturbation gives rise to sustained oscillations. \par
	
	The Rho activity oscillations that we predict agree in order of magnitude with earlier experiments \cite{Miller2009, Machacek2009} and theory \cite{Nikonova2013}. However, the period of the YAP oscillations that we predict at stationary $q$ is of the order of seconds, which differs considerably from the fluctuations on the scale of hours reported by Franklin et al. \cite{Franklin2020}. The oscillations in the latter study are induced by oscillating calcium levels, which affect many processes other than mechanotransduction, and the discrepancy in time scales suggests that they are due to a mechanism different from the one we propose. The period of the oscillations we observe is dictated by the rate of change of the gradient in active Rho, which is determined by the balance between the rates of Rho production and diffusion. The diffusion coefficients of the active and inactive forms of RhoGTPases are known \cite{Postma2004}, and the reasonable agreement between our results and the data in \cite{Miller2009,Machacek2009,Nikonova2013} suggests that the kinetic parameters we employ here are also of the correct order of magnitude. It follows, then, that the oscillation periods for the nuclear-to-cytoplasmic YAP ratio that we predict are realistic if the Rho-YAP coupling takes the form that we assume. We thus believe that it would be intriguing to conduct an experimental study of the time scales and region of parameter space that we consider here and that such a study is necessary to establish whether the different regimes we predict exist in biological systems. Note, however, that the time scales in our 1D model could potentially be significantly different from those in a real 3D system because the assumption that the diffusion coefficient values measured in such systems may not apply to our 1-D strip \cite{Zareh2012, Mirny2009}. Moreover, our one-dimensional system requires all species diffusing from one end of the cell to the other to pass through the nucleus, whereas in a real cell they could move around it instead. \par
	
	Finally, we find that applying a time-dependent mechanical stimulus $q$ to a system which for stationary $q$ qualitatively reproduces experimental data for $R$ versus adhesive area and only allows for positive YAP-Rho feedback results in a dependence of $R$ on time that closely follows that of $q$. \par
	
	Our model demonstrates that even a simplistic view of the chemical signalling pathway that governs cellular mechanotransduction can yield rich dynamic behaviour if mechanical stimuli are taken into account. The model qualitatively replicates existing experimental data for fixed values of the mechanical stimulus $q$ and predicts oscillatory regimes that would be intriguing to study experimentally. 
	This work is a first step in understanding how YAP activation and Rho signaling work together to enable cell to perceive its mechanical microenvironment. This is important because over the last decade YAP has emerged as the central signaling hub for mechanosensing and the Rho GTPase proteins have been established as key regulators of cytoskeletal dynamics. More importantly, mechanical activation of YAP through Rho signaling has been shown to play an important role in vital biological processes, from the expansion and survival of human embryonic stem cells \cite{Ohgushi2015} to viscoelastic feedback from the extracellular matrix that modulates proliferation and cancer metastasis \cite{Brusatin2018}. It is, therefore, essential to construct spatiotemporal models of YAP mechanotransdution that allow for testable predictions. The framework presented in this study can easily be extended to 2D, which would allow for exploring the impact of more complex mechanical stimulation such as anisotropy in mechanical cues acting on the cells and the corresponding spatial distribution of YAP within the cells. Furthermore, establishing the spatiotemporal dynamics of YAP expression within the cell and its interconnection with the cell's force-generating machinery is an important step towards building more cohesive models of cell motility and bridging the gap between mechanotransduction at the individual cell level and mechanical force generation by collective migration at the multicellular level.
	\par
	
	\section{Acknowledgments}
	
	This work has received funding from the European Union's Horizon 2020 research and innovation program under the Marie Sk\l odowska-Curie grant agreement No. 847523 'INTERACTIONS'. J.K.N., M.L.H. and M.H.J. acknowledge support from the StemPhys DNRF Center of Excellence (DNRF116). M.H.J. acknowledges support from the Danish Council for Independent Research, Natural Sciences (DFF-116481-1001). A.D. acknowledges support from the Novo Nordisk Foundation (grant No. NNF18SA0035142), Villum Fonden (Grant no. 29476), and the Danish Council for Independent Research, Natural Sciences (DFF-117155-1001). \par
	
	\bibliographystyle{rsc}
	\bibliography{YAP_arXiv}																					
			\newpage
			
			\break
			\pagenumbering{arabic}
			\appendix
			\counterwithin{figure}{section}
			\counterwithin{table}{section}
			\counterwithin{figure}{section}
			\counterwithin{equation}{section}
			
			\section{Supporting Information}
			
				The data used for generating the figures, as well as animations of the concentration profiles, raw concentration profile data, additional plots and MATLAB scripts that can be used to reproduce the data and figures are available online.
				Here we present some additional results and figures that complement the data in the main text.
				
				\subsection{Additional information about the model}				
				The harmonic potential we use is defined as
				\begin{equation}\label{eq7}		
				\phi(x) = 
				\begin{cases}
				\frac{\left(x-\frac{x_\mathrm{nucl \ front}+x_\mathrm{nucl \ back}}{2}\right)^2}{x_\mathrm{nucl \ back} -x_\mathrm{nucl \ front}}+\frac{1}{4}\left(x_\mathrm{nucl \ front} - x_\mathrm{nucl \ back}\right)
				& \text{if} \ t > 0 \text{ and } x_\mathrm{nucl \ front} \leq x \leq x_\mathrm{nucl \ back}\\
				0
				& \text{otherwise.}
				\\
				\end{cases}   	  			  	  
				\end{equation}
				
				We assume that the diffusion coefficient of active YAP varies between the cytoplasm and the nucleus as a hyperbolic tangent,
				\begin{equation}\label{eq12}
				D_{Y_\mathrm{act}} = D_{Y_\mathrm{nucl}} + \left(D_{Y_\mathrm{cyto}}-D_{Y_\mathrm{nucl}}\right)
				\left[1-\frac{1}{2}\left(\tanh{\left[F(x-x_\mathrm{nucl \ front})\right]}-\tanh{\left[F(x-x_\mathrm{nucl \ back})\right]}\right)\right],
				\end{equation}
				where $F = 30$ is a scaling factor chosen to yield a transition region considerably narrower than the cell length.

				\begin{table*}[htp]
					\small
					\caption{Parameter definitions, values and units not listed in Table~\ref{tbl:main_parameters}.}
					\label{tbl:additional_parameters}
					\makegapedcells 
					\begin{tabular*}{\textwidth}{@{\extracolsep{\fill}}lllll}
						\hline
						\makecell[l]{Parameter} & \makecell[l]{Meaning} & \makecell[l]{Value / range} & \makecell[l]{Units} \\
						\hline
						\makecell[l]{$c_5$} & \makecell[l]{constant governing the feedback of YAP on Rho} & \makecell[l]{$1-30$} & \makecell[l]{$\SI{}{s^{-1}} C_{0 \mathrm{A}}$}\\
						\hline
						\makecell[l]{$c_6$} & \makecell[l]{base activation rate for YAP} & \makecell[l]{1} & \makecell[l]{$\SI{}{s^{-1}}$} \\
						\hline
						\makecell[l]{$c_7$} & \makecell[l]{base deactivation rate for YAP} & \makecell[l]{1} & \makecell[l]{$\SI{}{s^{-1}}$} \\
						\hline
						\makecell[l]{$c_8$} & \makecell[l]{Hill function parameter for YAP} & \makecell[l]{$2.5-3.5$} & \makecell[l]{$C_{0 \mathrm{A}}^{-1}L$} \\
						\hline																			
						\makecell[l]{$c_9$} & \makecell[l]{Hill function parameter for YAP} & \makecell[l]{1} & \makecell[l]{$1$} \\
						\hline									
						\makecell[l]{$c_{10}$} & \makecell[l]{Hill function parameter for YAP} & \makecell[l]{5} & \makecell[l]{$C_{0 \mathrm{A}}^{-1}L$} \\
						\hline																					
						\makecell[l]{$q$} & \makecell[l]{effective charge of active YAP} & \makecell[l]{$0.042-42$} & \makecell[l]{$\SI{}{s^{-1}}C_{0 \mathrm{A}}^{-1}L^2$} \\
						\hline																					
						\makecell[l]{$q_{0}$} & \makecell[l]{effective charge of active YAP in the absence of a stimulus} & \makecell[l]{$1.25-2.5$} & \makecell[l]{$\SI{}{s^{-1}} L$} \\
						\hline
						\makecell[l]{$q_{1}$} & \makecell[l]{effective charge of inactive YAP} & \makecell[l]{$0.5-2$} & \makecell[l]{$\SI{}{s^{-1}}L$} \\
						\hline
						\makecell[l]{$Y_{\mathrm{act \ nucl \ threshold}}$} & \makecell[l]{threshold value above which YAP deactivates Rho} & \makecell[l]{$0.4-1$} & \makecell[l]{$N_{0 \mathrm{Y}}L^{-3}$} \\
						\hline			
						\makecell[l]{$x_\mathrm{nucl \ front}$} & \makecell[l]{$x$-coordinate of the front edge of the nucleus} & \makecell[l]{1/3} & \makecell[l]{$L$} \\
						\hline			
						\makecell[l]{$x_\mathrm{nucl \ back}$} & \makecell[l]{$x$-coordinate of the back edge of the nucleus} & \makecell[l]{2/3} & \makecell[l]{$L$} \\
						\hline			
						\makecell[l]{$A_\mathrm{pert}$} & \makecell[l]{amplitude of the initial parabolic perturbation} & \makecell[l]{$10^{-2}-1$} & \makecell[l]{$1$} \\
						\hline									
						\makecell[l]{$L_\mathrm{pert}$} & \makecell[l]{width of the perturbed region (for a parabolic initial perturbation)} & \makecell[l]{0.15} & \makecell[l]{$L$} \\
						\hline			
						\makecell[l]{$S_\mathrm{ampl}$} & \makecell[l]{amplitude of the transient stimulus taken from Ref.~\cite{Mori2008}} & \makecell[l]{$0.01-0.5$} & \makecell[l]{$\SI{}{s^{-1}} $} \\
						\hline						
						\makecell[l]{$t_1$} & \makecell[l]{duration of the application of a $t$-independent stimulus, see Ref.~\cite{Mori2008}} & \makecell[l]{20} & \makecell[l]{$\SI{}{s} $} \\
						\hline			
						\makecell[l]{$t_2$} & \makecell[l]{duration of the application of a $t$-dependent stimulus, see Ref.~\cite{Mori2008}} & \makecell[l]{25} & \makecell[l]{$\SI{}{s} $} \\
						\hline			
					\end{tabular*}
					\nomakegapedcells
				\end{table*}
				
				\begin{figure}
					\centering
					\subfloat[]
					{
						\includegraphics[width=8cm]{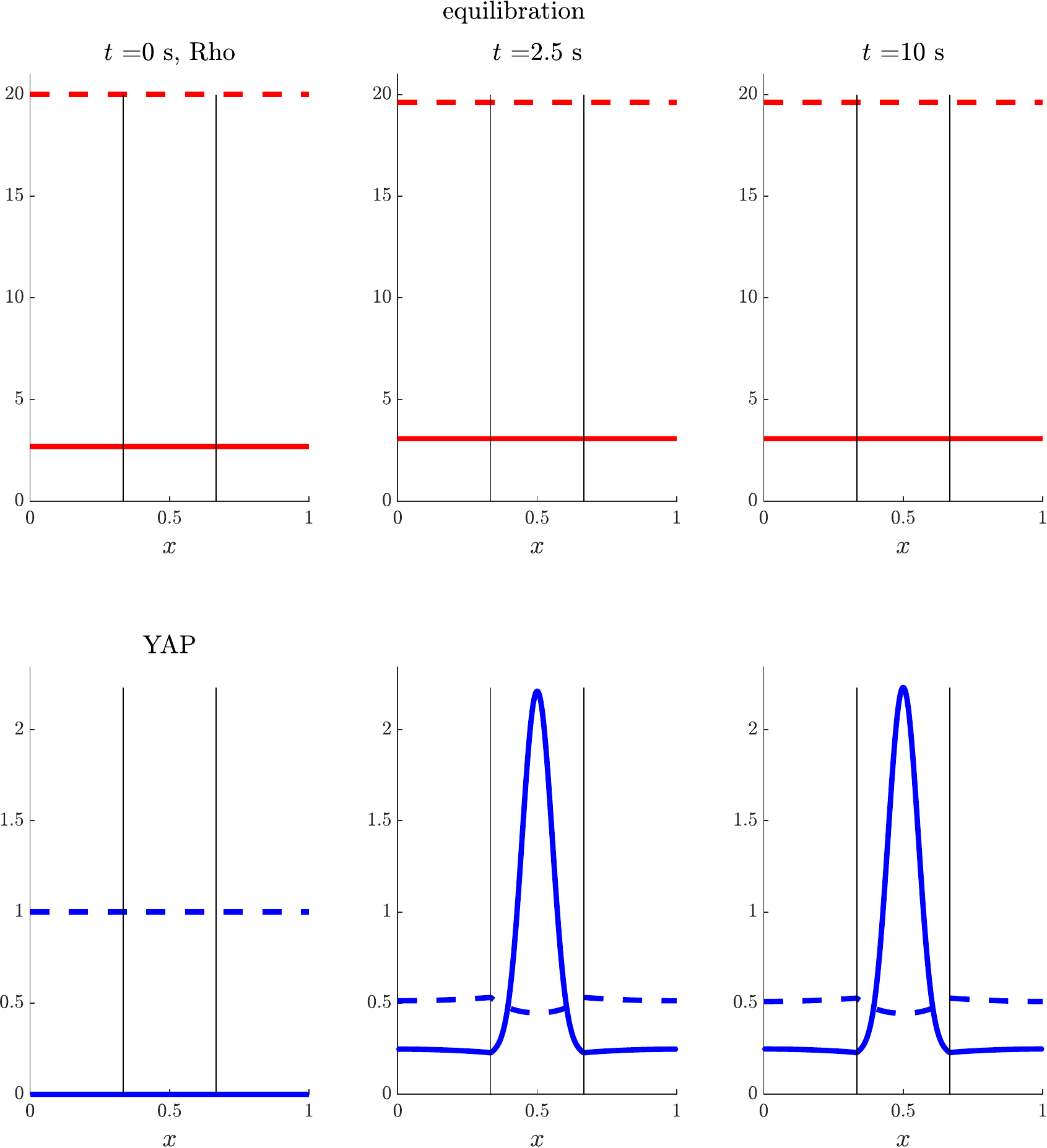}
					}
					\subfloat[]
					{
						\includegraphics[width=8cm]{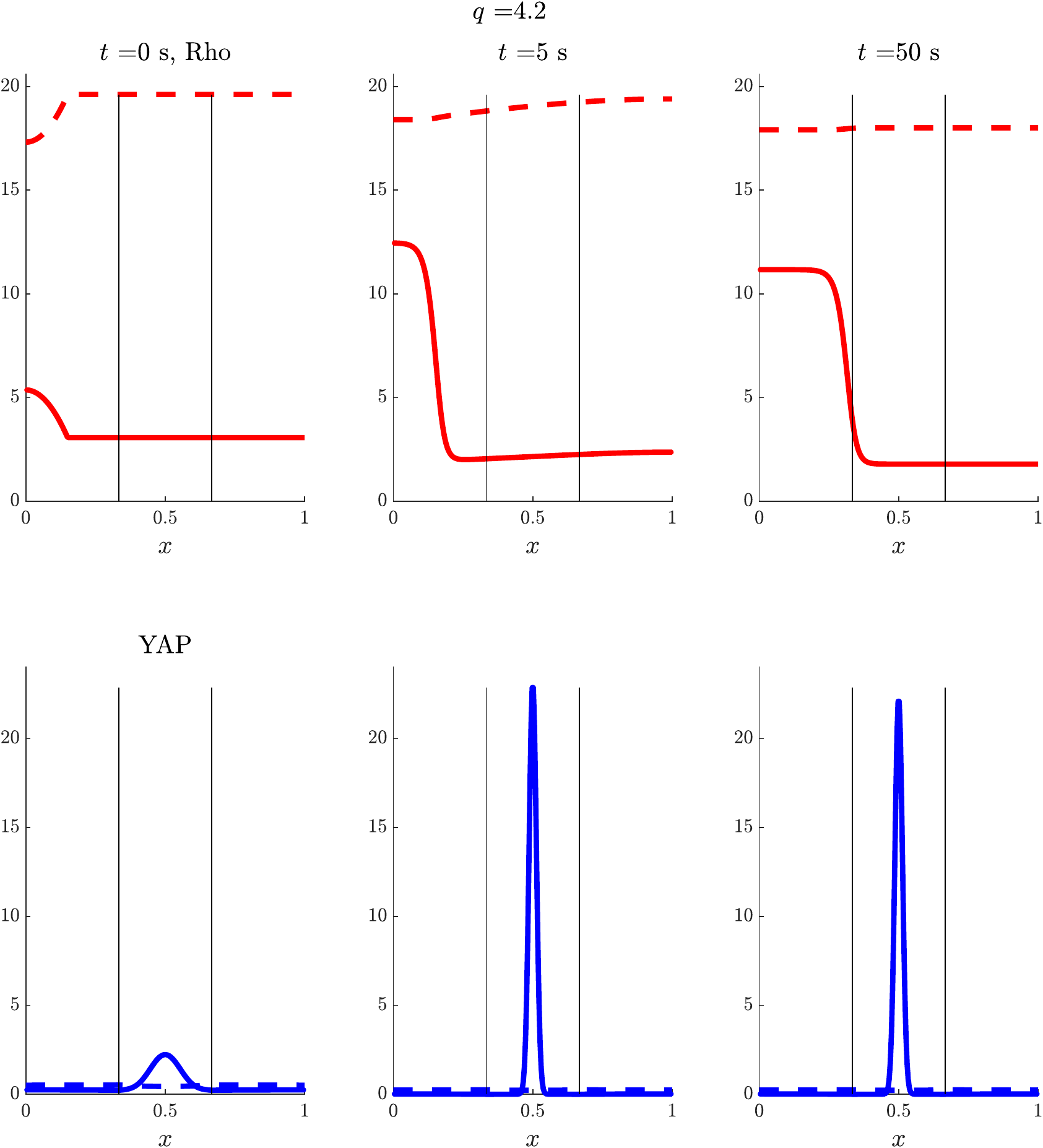}
					}
					\\
					\subfloat[]
					{
						\includegraphics[width=9cm]{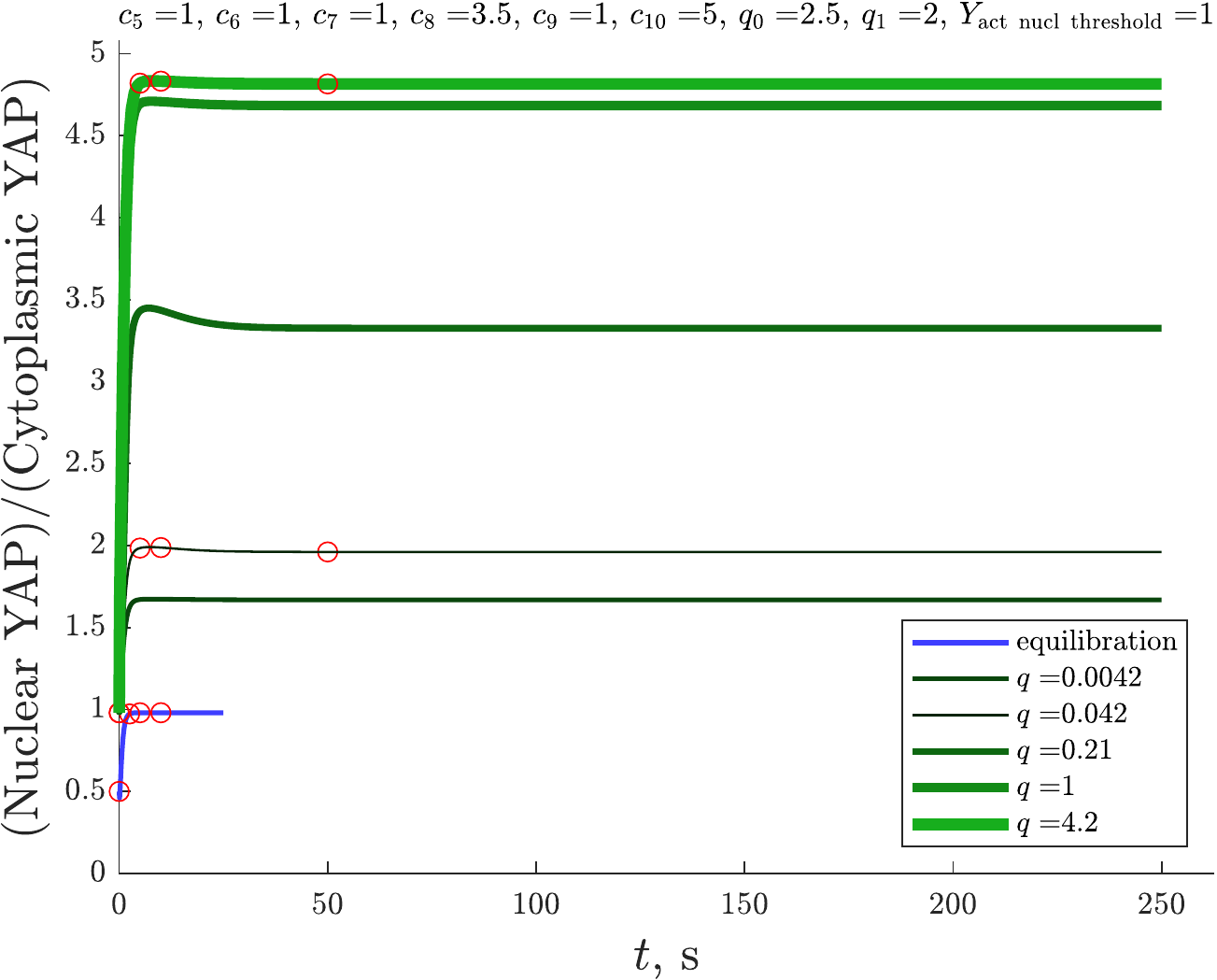}
					}
					\caption
					{
						\textbf{a}. Snapshots of the evolution of the concentrations of the different species in the system for an equilibration simulation with no perturbation in the Rho concentrations. Rho (top row) and YAP (bottom row), with solid lines depicting acting forms and dashed lines - inactive forms. The vertical lines indicate the boundaries of the nucleus.
						\textbf{b}. Snapshots of the evolution of the concentrations of the different species in the system at a high value of the mechanical forcing parameter $q$ - Rho (top row) and YAP (bottom row), with solid lines depicting acting forms and dashed lines - inactive forms. As seen in the plot for $t = \SI{0}{s}$, the initial perturbation to the homogeneous Rho concentrations is parabolic.  The vertical lines indicate the boundaries of the nucleus.
						\textbf{c}. Evolution of the YAP ratio for simulations with a parabolic initial perturbation at different values of the mechanical forcing parameter $q$. Red circles mark the points in time at which the concentration profile snapshots are taken.
					}
					\label{fgr:C_snapshots_equilibration_q=4_2_YAP_ratio_t_sweep}
				\end{figure}	
			
				\begin{figure}
					\centering
					\subfloat[]
					{
						\includegraphics[width=8cm]{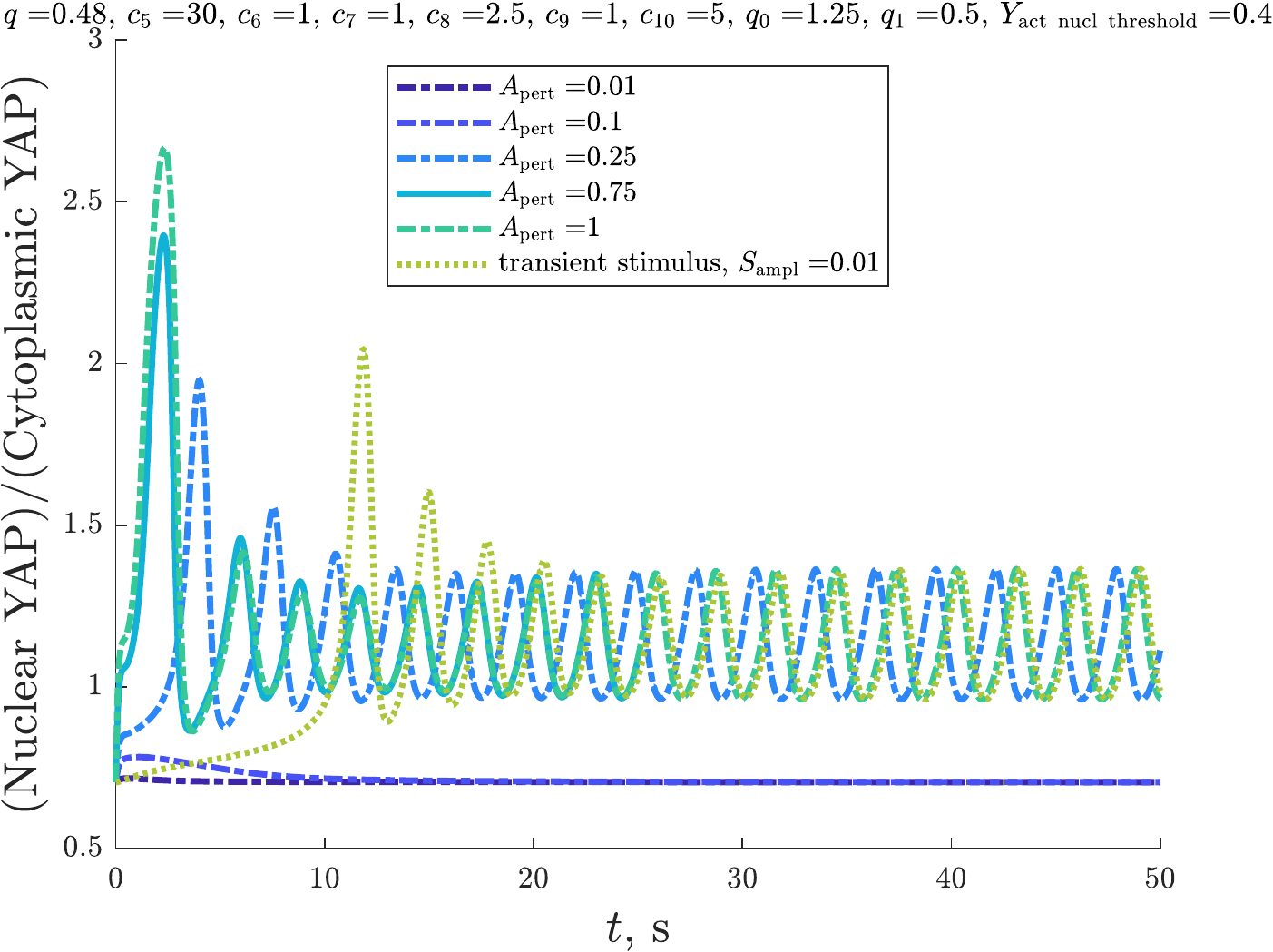}
					}
					\subfloat[]
					{
						\includegraphics[width=8cm]{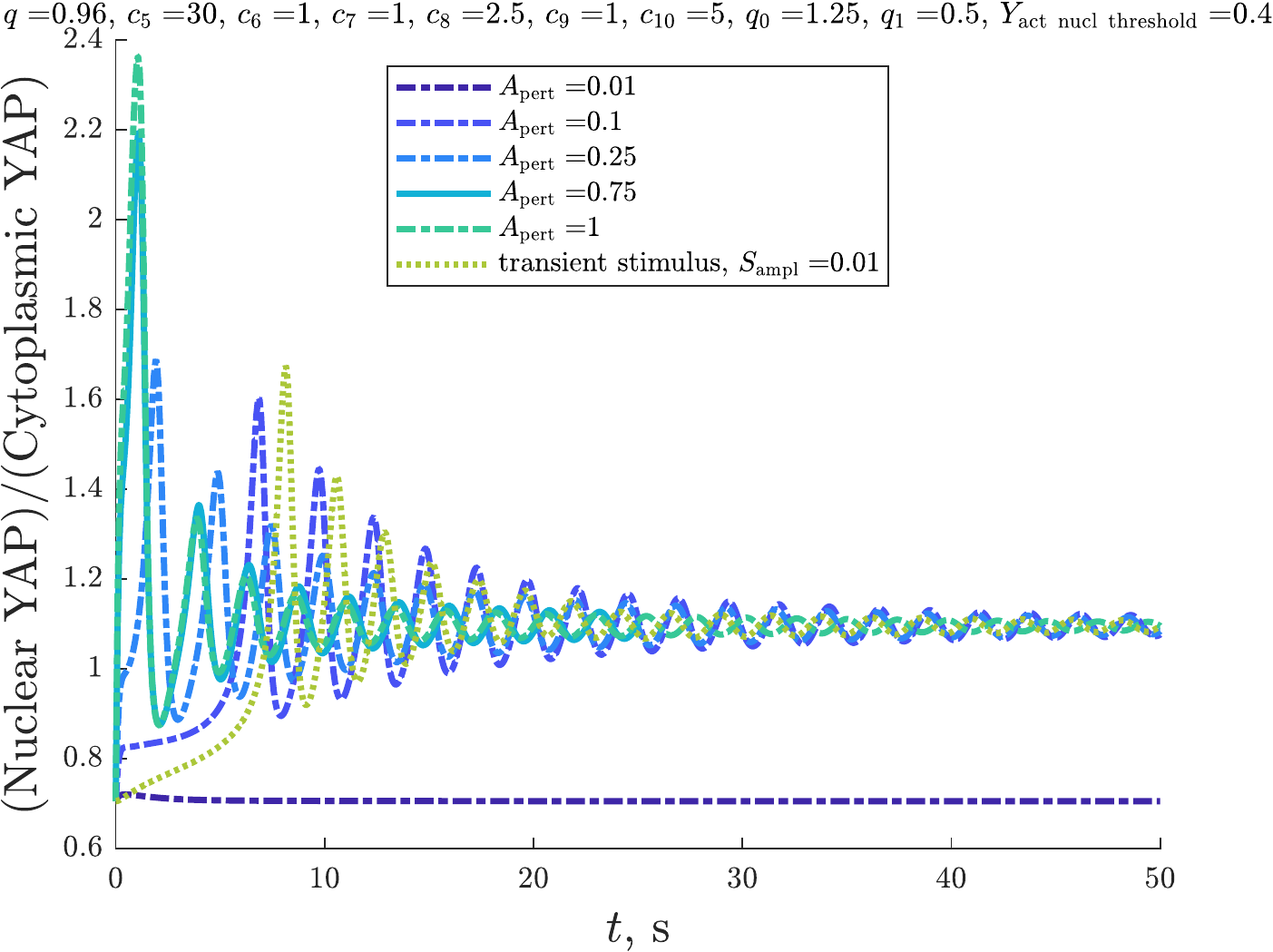}
					}
					\caption
					{
						$R(t)$ for simulations with different initial conditions, namely parabolic perturbations of varying amplitudes (eq.~\eqref{eq15}) and transient stimuli as per eqs.~\eqref{eq16}-\eqref{eq17}. The curves at $A_\mathrm{pert} = 0.75$ are those plotted in Figure~\ref{fgr:sust_damped_osc}a; \textbf{a} illustrates the case of $q = 0.48$, which allows for sustained oscillations, whereas at $q = 0.96$, only damped oscillations are possible. For the simulations with a transient stimulus as per Ref. \cite{Mori2008}, we use a much lower $S_\mathrm{Ampl}$ than the value of 0.5 in Figure~\ref{fgr:YAP_ratio_q_sweep_C_snapshots_q=0_042}a because the larger YAP-Rho coupling constant $c_5$ here would otherwise cause negative values of the concentrations of some species.
					}
					\label{fgr:R_t_diff_A_pert_trans_stim}
				\end{figure}
			
				\begin{figure}
					\centering
					\subfloat[]
					{
						\includegraphics[width=11cm]{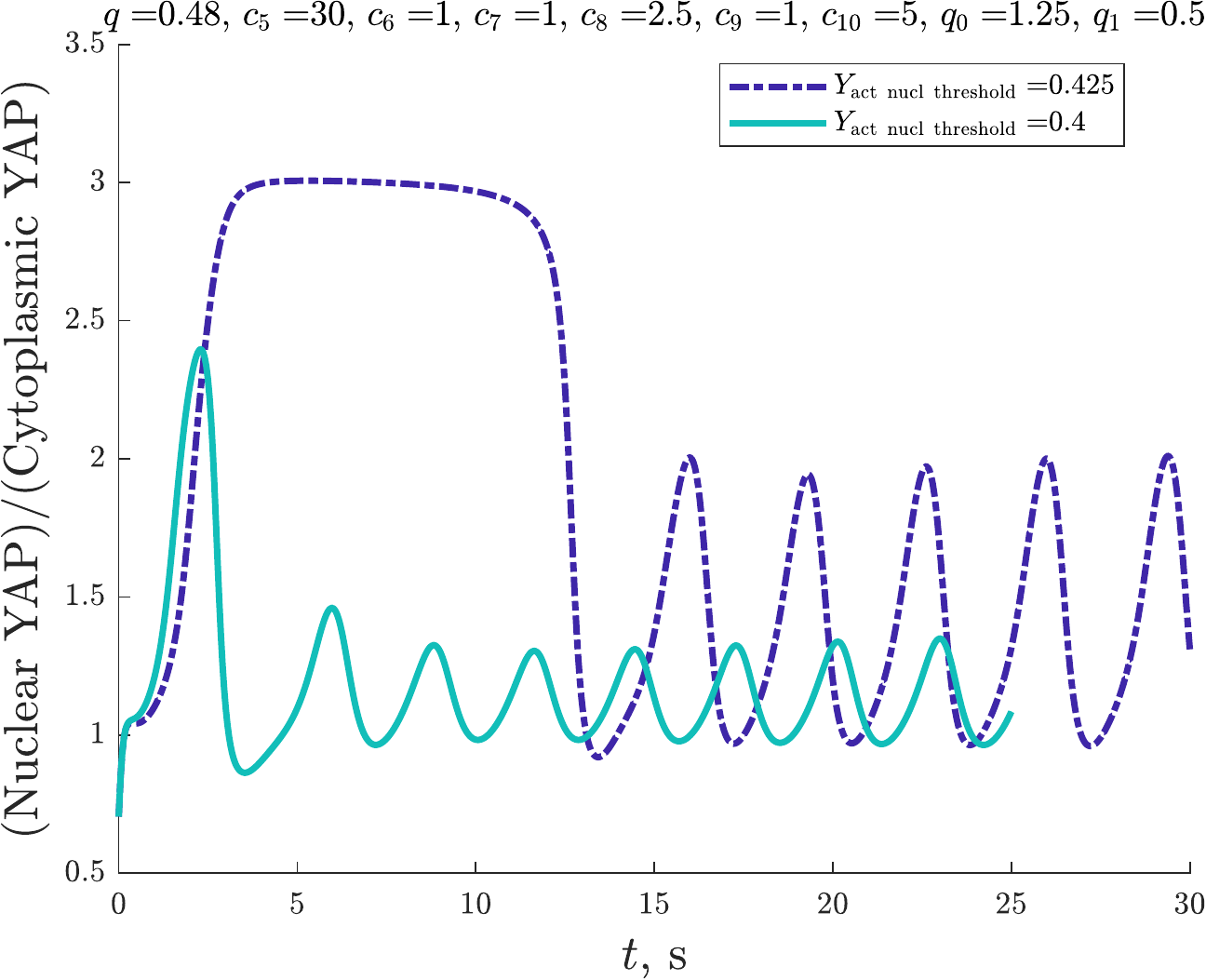}
					}
					\caption
					{
						$R(t)$ for a simulation with the same parameters as the one in Figure~\ref{fgr:sust_damped_osc}b except for $Y_{\mathrm{act \ nucl \ threshold}} = 0.425$ rather than $Y_{\mathrm{act \ nucl \ threshold}} = 0.4$. Note that $R(t)$ oscillates about $\approx 1.5$ rather than 1.15 as in Figure~\ref{fgr:sust_damped_osc}b.
					}
					\label{fgr:R_t_q_0_48_Yactnuclthr_0_425}
				\end{figure}
																			
\end{document}